\documentclass[superscriptaddress,twocolumn,amsmath,amssymb,prb,floatfix,citeautoscript,noshowpacs]{revtex4}
\usepackage{graphicx}
\usepackage{dcolumn}
\usepackage{bm}
\usepackage{color}
\usepackage{ulem}
\usepackage{tikz}
\usepackage{rotating}
\usepackage{amsmath}
\usepackage{amssymb}

\begin{document}
\title{Spin Vortices and Skyrmions of a Single Electron in Inhomogeneous Magnetic Fields}
\author{Amin Naseri}
\affiliation{School of Physics and Electronics, Central South University, Changsha, P. R. China 410083}

\affiliation{Department of Physics and Astronomy, University of Manitoba, Winnipeg,
Canada R3T 2N2}

\author{Shenglin Peng}
\affiliation{School of Physics and Electronics, Central South University, Changsha, P. R. China 410083}

\author{Wenchen Luo}
\email{luo.wenchen@csu.edu.cn}
\affiliation{School of Physics and Electronics, Central South University, Changsha, P. R. China 410083}

\author{Jesko Sirker}
\email{sirker@physics.umanitoba.ca}
\affiliation{Department of Physics and Astronomy, University of Manitoba, Winnipeg,
Canada R3T 2N2}
\date{\today}

\begin{abstract}
We study the spin textures of a confined two-dimensional electron in
inhomogeneous magnetic fields. These fields can either be external or
effective fields due to a background magnetic texture in the plane in
which the electron resides. By analytical considerations, WKB-type
approximations, and by performing numerical diagonalizations we show
that the in-plane spin field components of a single electron can form
vortices while the total spin field can become a skyrmion. Most
interestingly, we find that topological trivial magnetic fields can
induce topological spin field configurations in the eigenstates of the
electron due to quantum effects.
\end{abstract}

\maketitle

\section{Introduction}
Pursuing topological phenomena has become one of the main themes in
condensed matter physics \cite{Nat547}. In this context, topology
refers to the existence of an integer quantity (the topological charge)
which is related to the global properties of the system and
insensitive to small local perturbations \cite{thoul}. In the approach
introduced by Michael Berry \cite{berry01}, adiabatically varying the
parameters of a model generates an effective magnetic field in its
parameter space
\cite{berry02}. The field is induced by topological charges
in close analogy with Dirac monopoles. The Berry phase, which is the
observable effect of a topological charge, can be understood as the
flux of the emergent field through a surface in parameter space. We
demonstrate below that from the Schr\"odinger equation describing a
two-dimensional (2D) electron without a good spin quantum number,
current densities can be derived which can be directly related to the
topological charge of the spin texture.

The spin of a single electron confined in a quantum dot \cite{dot01,dot02} realizes a
quantum bit (qubit) which can encode information \cite{PRA57,Nat489}
and can be used as a building block for a quantum computer
\cite{Zutic,Smejkal,Sinova}. Apart from potential applications, recent
experimental progress in controlling a single electron and its
confining potential \cite{science339,science318,PRL122,PRB99} also
raises fundamental questions about the spin texture of single
electrons in internal effective and applied external magnetic
fields. Our study is motivated by the following specific questions:
How do the spin fields of a single electron arrange themselves in the
presence of a magnetic field with a trivial or non-trivial topological pattern?
In a recent work, Ref.~\onlinecite{SOCST}, it has been reported that
the in-plane spin texture of the ground state of a 2D electron in a
spin-orbit coupled quantum dot can develop a vortex with a topological
charge $|q|=1$. Furthermore, the interplay of spin and the orbital
angular momenta of electron vortex beams \cite{elvo01,elvo02}
propagating in a 3D space subject to electric and magnetic fields has
been studied theoretically and experimentally \cite{elvo03,elvo04}. In
a setting more relevant to our study, electromagnetic fields with
topological patterns have been used to directly manipulate the spin of
electron beams \cite{elvo05}.

In this work, the spin texture of an electron strongly confined to
a 2D surface is explored in the presence of an inhomogeneous magnetic
field $\mathbf{B}=(B_{1},B_{2},B_{3})$ \cite{Nogaret}. The magnetic
field $\mathbf{B}$ can represent an external field or the effects
of the local magnetic moments of the material which hosts the electron
\cite{Kondo}. In the main text, the focus is on the spin textures of
eigenfunctions of the 2D system in the presence of spatially dependent
inhomogeneous fields. We analyze these textures both analytically and
numerically. For magnetic fields which themselves carry a topological
charge, the semiclassical Landau-Lifshitz-Gilbert (LLG) equations turn
out to be sufficient for this purpose. The spin texture then simply
follows the magnetic field---in a similar way small magnetic needles
follow the magnetic field lines---and thus also acquires a non-zero
topological charge. If the Hamiltonian has a certain rotational
symmetry then it is even possible to establish an exact relation
between the in-plane spin texture and the topological magnetic
field. This relation is in agreement with the LLG equations. More
surprisingly, we find that also topologically trivial magnetic fields
can induce topological spin textures. In the latter case, it is
paramount to consider the interplay between the Zeeman energy and the
kinetic energy of the electron. We show that the topological spin
patterns which can develop in this case are understood using a
Wentzel-Kramers-Brillouin (WKB) approximation.

We consider two classes of topologically trivial magnetic fields: On
the one hand, continuous fields which can be thought of as being
projections of fields with a topological pattern onto a plane. On the
other hand, discontinuous effective fields which might arise due to
magnetic domain walls within the plane the electron is confined to. We
find that if $\mathbf{B}$ is a trivial field but induces a non-trivial
spin texture, then different eigenstates can have different
topological numbers. This is in contrast to the case when $\mathbf{B}$
itself has a definite topological pattern in which case all
eigenstates have the same topological number as the field. We want to
stress that not all of the magnetic fields considered here can be
realized as external fields. They rather represent effective magnetic
fields originating from the magnetic texture of the background plane
which is Hund-coupled
\cite{Skyrmion} with the spin of the electron. In addition, we will also
analytically address the possible spin textures due to a uniform
magnetic field and internal spin-orbit couplings (see
Appendix~\ref{APPSOC}). Although our analysis is primarily concerned
with the spin texture of a single electron, it is applicable to any
spin-$1/2$ particle.

The manuscript is organized as follows. In Sec.~\ref{SecPre}, we
introduce a generic 2D Hamiltonian of an electron in an inhomogeneous
magnetic field. Continuity equations for the spin densities and
current densities for a generic eigenfunction are presented. This
representation also helps to establish identities which connect the
topological charges to the charge density, the spin densities, and
their current densities. In Sec.~\ref{SecDeC}, we discuss the spin
textures generated by magnetic fields with a topological charge. The
possible spin textures in topologically trivial magnetic fields are
presented in Sec.~\ref{SecMFWTTP}. The two different cases, continuous
and discontinuous fields, are analyzed in separate subsections. In
Sec. \ref{SecNoGolTo}, magnetic fields with different winding numbers
in different domains are discussed. A brief summary of our main
results and an outlook is given in Sec.~\ref{Concl}.

\section{Preliminaries}
\label{SecPre}

The Hamiltonian $H$ of a non-relativistic electron
in a magnetic field is given by
\begin{equation}
\label{eq_H}
H=H_{0}+\frac{\Delta}{2}\mathbf{B}\cdot\boldsymbol{\sigma},
\end{equation}
where $\Delta$ is a Zeeman coupling factor, $\boldsymbol{\sigma}=(\sigma_{1},\sigma_{2},\sigma_{3})$ is the vector of Pauli matrices, and
\begin{eqnarray}
H_{0}=\frac{\boldsymbol{\Pi}^{2}}{2m_{e}}+V_{0} \, .
\label{eq_H0}
\end{eqnarray}
$H_{0}$ commutes with all Pauli matrices, $[H_{0},\sigma_{j}]=0$ for
$j=1,2,3$, $\boldsymbol{\Pi}=\mathbf{p}-e\mathbf{A}$ is the mechanical
momentum, $\mathbf{A}$ is a vector potential and we use the Coulomb
gauge $\nabla\cdot\mathbf{A}=0$. $V_0$ is a confining potential. The
mass and charge of the electron are $m_{e}$ and $e<0$, respectively.
In numerical calculations, we use the parameters appropriate for an
InAs quantum dot: $m_{e}=0.042 m_0$ where $m_0$ is the mass of a free
electron, and a Land\'e factor $g_{e}=-14$.  The two independent
energy scales of the theory are the cyclotron energy
$\hbar\omega_{c}=\hbar|e|B_{0}/m$ and the Zeeman energy $\Delta
B_{0}$, where $B_{0}$ is the unit of magnetic field in our study. Here
we will investigate the spin textures of an electron in the quantum
regime, in which the eigenfunctions of $H$ govern the physics of the
electron.

If $\Psi(\mathbf{r})$ is an eigenfunction of $H$, then the charge and spin
fields are defined by
\begin{eqnarray}
\varrho_{\nu}\equiv
\Psi^{\dag}(\mathbf{r})\sigma_{\nu}\Psi^ {}(\mathbf{r}),
\end{eqnarray}
for $\nu=0,1,2,3$ where $\sigma_{0}$ is the $2\times2$ identity
matrix, and the position vector---in the plane the electron resides
in---is given by is $\mathbf{r}=(\rho,\theta)$. Occasionally, we refer
to the fact that the 2D system is embedded in 3D space, i.e., the
electron is living at $z=0$ in a cylindrical coordinate system
$\mathbf{r}=(\rho,\theta,z)$. Investigating the non-trivial topology
of the in-plane spin field
\begin{eqnarray}
\boldsymbol{\varrho}_{\bot}(\mathbf{r})=(\varrho_{1},\varrho_{2}),
\end{eqnarray}
and the full three-dimensional spin field
\begin{eqnarray}
\boldsymbol{\varrho}(\mathbf{r})=(\varrho_{1},\varrho_{2},\varrho_{3}),
\end{eqnarray}
is the purpose of this work.
We note that $\boldsymbol{\varrho}(\mathbf{r})$ and
$\boldsymbol{\varrho}_{\bot}(\mathbf{r})$ are observables and a unitary
transformation of $H$ leaves their topological features invariant
\cite{Nakahara}.

We review two simple cases which do not result in a topological
pattern in the spin texture. First, if the Hamiltonian is
translationally invariant $[H,\mathbf{p}]=0$, where
$\mathbf{p}=-i\hbar\nabla$ is the canonical momentum, then $\Psi$ can
be chosen to be a simultaneous eigenfunction of $H$ and $\mathbf{p}$
which leads to $\mathbf{p}\varrho_{j}=0$ since
$\left(\mathbf{p}\Psi\right)^{\dag}=-\mathbf{p}\Psi^{\dag}$
(i.e. $\Psi^ {}$ and $\Psi^{\dag}$ have opposite momenta). Second, if
there is a good spin quantum number, then by a unitary transformation
$\sigma_{3}$ can be chosen to commute with the Hamiltonian
$[H,\sigma_{3}]=0$, therefore, $\varrho_{1,2}=0$ and
$\varrho_{3}=\pm\varrho_{0}$. We conclude that if the spin texture is
non-trivial, then the translational symmetry must be broken and
$[H,\boldsymbol{\sigma}\cdot\hat{n}]\neq0$ must be satisfied for any
fixed 3D unit vector $\hat{n}$.

In the following, the translational symmetry is broken by the
inhomogeneous magnetic field and by the confining scalar potential
$V_{0}$. The symmetries and the analytical form of the confining
potential do not play a role in our results. We assume that the
electron is strongly confined to the 2D plane by a potential
$m\omega_{z}^{2}z^{2}/2$ which allows to neglect the degree of freedom
perpendicular to the plane. Within the plane, we assume that the
electron has confinement lengths $L_{x,y}$.

An eigenfunction of $H$, which is a spinor, can be formally written
as
\begin{eqnarray}
\Psi(\mathbf{r})=\begin{pmatrix}e^{iS_{1}}\psi_{1}\\
e^{iS_{2}}\psi_{2}
\end{pmatrix},\label{eq_GEF}
\end{eqnarray}
where $S_{1,2}$ and $\psi_{1,2}$ are real functions. The charge
and spin densities become
\begin{eqnarray}
\varrho_{0}&=&\psi_{1}^{2}+\psi_{2}^{2},\quad \varrho_{1}=2\psi_{1}\psi_{2}\cos{\left(S_{2}-S_{1}\right)},\nonumber \\
\varrho_{2}&=&2\psi_{1}\psi_{2}\sin{\left(S_{2}-S_{1}\right)},\quad \varrho_{3}=\psi_{1}^{2}-\psi_{2}^{2},
\end{eqnarray}
where $\varrho_{0}=|\Psi^ {}(\mathbf{r})|^{2}$. For the 3D vector
field, it follows immediately that
\begin{eqnarray}
\sum_{j=1}^{3}\varrho_{j}^{2}=\varrho_{0}^{2}, \label{eqNorFi01}
\end{eqnarray}
which implies that the spin is conserved globally
\begin{eqnarray}
\int d^{2}r|\boldsymbol{\varrho}(\mathbf{r})|=\int d^{2}r\varrho_{0}=1.\label{eqSCG01}
\end{eqnarray}
Using the representation \eqref{eq_GEF}, we find that the phase difference
between the two spin components is related to the in-plane spin fields
\begin{eqnarray}
\nabla(S_{2}-S_{1})=\frac{\varrho_{1}\nabla\varrho_{2}-\varrho_{2}\nabla\varrho_{1}}{\varrho_{1}^{2}+\varrho_{2}^{2}}.
\end{eqnarray}

The {\it winding number of the in-plane spin field} is now obtained by requiring that the eigenfunction is single-valued,
\begin{eqnarray}
q=\frac{1}{2\pi}\oint\nabla\left(S_{2}-S_{1}\right)\cdot d\mathbf{l},
\,\,\,\,\,\,\,\,q\in\mathbb{Z},\label{Def_q01}
\end{eqnarray}
where $d\mathbf{l}$ is an element of a closed loop in the $x-y$
plane. If $q=0$, then the spin field is trivial. If $q\neq 0$, then
$\boldsymbol{\varrho}_{\bot}(\mathbf{r})$ describes a vortex with the
topological charge $q$. If the radius of the loop $\rho$ is taken such
that $\rho\gg L_{x,y}$ and the loop encloses all singularities, then
we call $q$ the net winding number of the spin
configuration. We can understand the relation between
the winding number and the Berry connection of the wave function in
the special case where the kinetic energy can be ignored and
$B_3=0$. In this case $\varrho_3=0$ and the winding number can be
written immediately in terms of a Berry connection
\begin{eqnarray}
A_{B}=i\langle\Psi_{B}|\nabla\Psi_{B}\rangle=\nabla\left(S_{1}-S_{2}\right),\label{eq_connection01}
\end{eqnarray}
where we have fixed the gauge by setting
\begin{eqnarray}
|\Psi_{B}\rangle=\frac{1}{\sqrt{2}}\begin{pmatrix}1\\
e^{i(S_{2}-S_{1})}
\end{pmatrix}.
\end{eqnarray}

On the other hand, we can also consider the topological character of
the full three-dimensional spin field $\boldsymbol{\varrho}(\mathbf{r})$. To do so,
we define the topological current density
\begin{eqnarray}
J_{t} & = & \boldsymbol{\bar{\varrho}}(\mathbf{r})\cdot\left(\partial_{x}\boldsymbol{\bar{\varrho}}(\mathbf{r})\times\partial_{y}\boldsymbol{\bar{\varrho}}(\mathbf{r})\right),
%
%
\end{eqnarray}
where $\boldsymbol{\bar{\varrho}}(\mathbf{r})=\boldsymbol{\varrho}(\mathbf{r})/\varrho_{0}$. The
{\it Skyrmion (Pontryagin) number} of the spin field can then be expressed as
\begin{eqnarray}
Q=\frac{1}{4\pi}\int d^{2}r \, J_{t}.
\label{Skyrmion}
\end{eqnarray}
In this article we will investigate both the in-plane winding number
$q$ defined in Eq.~\eqref{Def_q01} as well as the Skyrmion number $Q$
which are different in general.

If we know the eigenfunctions $\Psi$ of the Hamiltonian, for example
from a numerical calculation, then the spin fields and their
topological numbers can be easily calculated. Here we will show that
it is possible to understand the topology of the spin textures
analytically without having to fully diagonalize the Hamiltonian. In
order to do so, we will start from continuity equations for the spin
fields \cite{spcu01,spcu02} which are decompositions of the charge
continuity equation in the basis of Pauli matrices. Multiplying the
time-dependent Schr\"odinger equation from the left by
$\Psi^{\dag}\sigma_{\nu}$ and taking its imaginary part, we find that
the spin current densities
\begin{eqnarray}
\mathbf{J}_{\nu}=\frac{1}{2m_{e}}\left[\Psi^{\dag}\sigma_{\nu}\mathbf{p}\Psi-\left(\mathbf{p}\Psi^{\dag}\right)\sigma_{\nu}\Psi\right]-\frac{e}{m_{e}}\mathbf{A}\varrho_{\nu},\label{eqCurrDen01}
\end{eqnarray}
obey the continuity equations
\begin{eqnarray}
\nabla\cdot\mathbf{J}_{\nu}+\partial_{t}\varrho_{\nu}=\frac{\Delta}{\hbar}\left[\mathbf{B}\times\boldsymbol{\varrho}(\mathbf{r})\right]\cdot\hat{x}_{\nu},\label{eqSCE01}
\end{eqnarray}
where $\hat{x}_{j}$ are unit vectors in a 3D Cartesian coordinate
system, and we define $\hat{x}_{0}=0$. For a stationary system, the
time derivative of the densities is zero in all the equations,
$\partial_{t}\varrho_{\nu}=0$. If there is a good spin quantum number
e.g. $[H,\sigma_{3}]=0$, all the terms in the continuity equations for
$\varrho_{1,2}$ vanish while the continuity equation for the third
component of the spin density reduces to the charge density continuity
equation $\nabla\cdot\mathbf{J}_{0}+\partial_{t}\varrho_{0}=0$.  If
the spin is not a good quantum number, then the components of the spin
torque play the role of a source/sink in the continuity equations of
the spin fields. Since the spin is conserved globally according to
Eq.~\eqref{eqSCG01}, the net contribution of the spin-torque
components must vanish. This can be verified as follows
\begin{eqnarray}
& & i\Delta \int d^{2}r\,\mathbf{B}\times\boldsymbol{\varrho}(\mathbf{r})  = \langle\Psi|\left[\boldsymbol{\sigma},H\right]|\Psi\rangle \\
&=& \langle\Psi|\left(\boldsymbol{\sigma}H-H\boldsymbol{\sigma}\right)|\Psi\rangle =  \langle\Psi|\boldsymbol{\sigma}|\Psi\rangle\left(E-E\right)=0 \nonumber \, ,
\end{eqnarray}
where $E$ is an eigenenergy of the Hamiltonian, $H|\Psi\rangle=E|\Psi\rangle$.
This relation implies $\int d^{2}r\,\nabla\cdot\mathbf{J}_{\nu}=0$,
by taking an integral of Eq.~\eqref{eqSCE01} in the static case.

The topological numbers $q$ and $Q$, defined in Eq.~\eqref{Def_q01}
and Eq.~\eqref{Skyrmion} respectively, can be rewritten in terms of
the current densities, see Appendix \ref{AppWNCD} for details. The
winding number $q$ is related to $\mathbf{J}_{0}$ and $\mathbf{J}_{3}$
by
\begin{eqnarray}
q=\frac{m_{e}}{\pi\hbar}\oint\frac{\varrho_{3}\mathbf{J}_{0}-\varrho_{0}\mathbf{J}_{3}}{\varrho_{0}^{2}-\varrho_{3}^{2}}\cdot d\mathbf{l},\label{Def_q02}
\end{eqnarray}
and the topological current density has the form (with a sum over repeated indices implied)
\begin{eqnarray}
J_{t}=\frac{\varrho_{\nu}}{\varrho_{0}}\eta^{\nu\mu}\frac{b_{\mu}}{\varrho_{0}},\label{eq_Q_02}
\end{eqnarray}
where
$b_{\mu}=\frac{m_{e}}{\hbar}\nabla\times\mathbf{J}_{\mu}\cdot\hat{z}$,
and $\eta^{\nu\mu}$ is the Minkowski metric with a signature
$(+,-,-,-)$.  It is worth mentioning that the representation of
$J_{t}$ in Eq. (\ref{eq_Q_02}) is the Mermin-Ho relation
\cite{Skyrmion} for the case $\varrho_{0}\neq1$.  These relations
suggest that the current densities play the role of vector potentials
for the emergent magnetic fields and the spatial integral of the
generated flux is the geometrical phase of the theory, see Appendix \ref{AppTDCD} for details.

\section{Magnetic fields with a topological pattern}
\label{SecDeC}

In this section, we study magnetic fields which themselves do have a non-zero winding or Skyrmion number defined by
\begin{eqnarray}
q_{B}=\frac{1}{2\pi}\oint\frac{\left({B_{1}}\nabla{B_{2}}-{B_{2}}\nabla{B_{1}}\right)}{B_{1}^{2}+B_{2}^{2}}\cdot d\mathbf{l}.\label{eqPhaB03}
\end{eqnarray}
and
\begin{eqnarray}
Q_{B}=\frac{1}{4\pi}\int d^{2}r\,\bar{\mathbf{B}}\cdot\left(\partial_{x}\bar{\mathbf{B}}\times\partial_{y}\bar{\mathbf{B}}\right),
\end{eqnarray}
respectively, where $\mathbf{\bar{B}}=\mathbf{B}/|\mathbf{B}|$. 
In order to find a relation between the spin densities and the components
of the magnetic field, we start from the continuity equations. First,
we note that if the Hamiltonian does not have a good spin quantum
number but commutes with the following angular-momentum operator
\begin{eqnarray}
	\jmath^{N}_{z}
	=
	\frac{-i}{N}\partial_{\theta}
	+
	\frac{1}{2} \sigma_{z},
	\quad
	N \in \mathbb{Z} \text{ and } N \neq 0,
	\label{eq_RotOp}
\end{eqnarray}
then the radial and angular degrees of freedom of the eigenfunction
$\Psi$ are separable and it can be shown that $\nabla \cdot
\mathbf{J}_{3}=0$ (see Appendix \ref{AppExRe}) which gives
\begin{eqnarray}
\varrho_{2} B_{1} =\varrho_{1} B_{2}.\label{eqPhaExact}
\end{eqnarray}
That is, the in-plane spin texture $\boldsymbol{\varrho}_{\bot}$
exactly follows the in-plane magnetic field $\mathbf{B}_{\bot}$, and
hosts a vortex with $q=N$. Therefore, it is proper to call the integer
number $N$ the vorticity of the rotational symmetry. Among magnetic
fields discussed below, the fields in Eqs.~(\ref{eq_Bn},\ref{eq:Bp}), and
\eqref{eq_Skyrm} used in the Hamiltonian \eqref{eq_H} have this symmetry. 
Hence, the vorticity of the fields in Eqs.~\eqref{eq_Bn},
~\eqref{eq:Bp} and ~\eqref{eq_Skyrm} are $N=-n$, $N=n$ for
$n\in\mathbb{Z}^{+}$ and $N=1$, respectively.  

We will now show that the relation \eqref{eqPhaExact} remains valid
approximately even if $\jmath^{N}_{z}$ is not conserved. To do so, we
consider the continuity equations at distances far from the origin of
the confinement, $\rho\gg L_{x,y}$. In this regime, the bare kinetic
energy is much smaller than the potential energies and can be
neglected. We further assume an in-plane vector potential 
$\mathbf{A}=B_{0}/2(-y,x,0)$ which gives a uniform magnetic field
perpendicular to the plane. The continuity equations \eqref{eqSCE01}
then reduce to
\begin{eqnarray}
-\frac{e}{m_{e}}\mathbf{A}\cdot\nabla\varrho_{j}
=\omega_{c}\partial_{\theta}\varrho_{j}\approx\frac{\Delta}{\hbar}\left[\mathbf{B}\times\boldsymbol{\varrho}(\mathbf{r})\right]\cdot\hat{x}_{j},\label{eqSCE02}
\end{eqnarray}
where $\omega_{c}=|e|B_{0}/(2m_{e})$ is the cyclotron frequency. We,
furthermore, define the dimensionless $\bar{\omega}_{c}=\hbar
\omega_{c}/E$ and a dimensionless magnetic field $\tilde{\mathbf{B}}=\Delta \mathbf{B}/E$
where $E$ is the eigenenergy of the eigenstate $\Psi$ we measure the
spin fields in. We then treat $\bar{\omega}_{c}$ as the smallest
parameter in the theory and expand the spin fields in a power series
in $\bar{\omega}_{c}$ in the spirit of the WKB approximation
\begin{eqnarray}
\label{spinexpand}
\varrho_{j}=\sum_{n=0}^\infty\bar{\omega}_{c}^{n}\varrho_{j}^{(n)}.
\end{eqnarray}
By setting equal the terms of the same power in $\bar{\omega}_{c}$
in Eq.~(\ref{eqSCE02}), the leading-order spin fields are related
to the magnetic field by
\begin{eqnarray}
\frac{\varrho_{i}^{0}}{\varrho_{j}^{0}}=\frac{B_{i}}{B_{j}},\label{eqPhaB02}
\end{eqnarray}
given that neither of the components of the magnetic field is
identically zero, $B_{j}\neq 0$. In other words, the spin fields to
leading order are aligned with the magnetic field in agreement with
what is expected classically and the exact relation in
Eq.~\eqref{eqPhaExact} in the rotationally symmetric case.
Given that $q_{B}$ is unique for all the contours enclosing the origin
(and not just those at $\rho\gg L_{x,y}$), the winding number of the
induced spin texture is given by $q=q_{B}$. It is worth noting that
all the eigenstates of $H$ have the same net winding number because
the above argument is applicable to any generic eigenstate of
$H$. However, Eq.~\eqref{eqPhaB02} is unable to predict the Skyrmion
charge of the spin texture even for a magnetic field $\mathbf{B}$ with
$Q_{B}\neq0$. Eq.~\eqref{eqPhaB02} is only valid at $\rho\gg L_{x,y}$
while computing the Skyrmion charge requires to know the polarisation
of $\varrho_{3}$ at every point of the plane.



\begin{figure}[tbh]
\begin{centering}
\begin{tikzpicture}

\begin{scope}[xshift=0.00\textwidth,yshift=0.23\textwidth]
\node[anchor=south west,inner sep=0](image) at (0,0){
\includegraphics[width=0.25\textwidth]{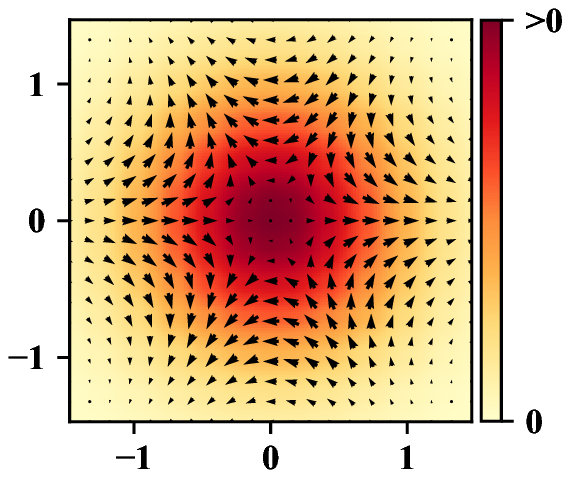}};
 \begin{scope}[x={(image.south east)},y={(image.north west)}]
 \node at (0.06,0.89) { $\mathbf{\left.{a}\right)}$};
 \node at (0.48,-0.02) {$\bm{x/l}$};
 \node at (0.04,0.45) {\begin{rotate}{90}$\bm{y/l}$\end{rotate}};
 \end{scope}
\end{scope}

\begin{scope}[xshift=0.25\textwidth,yshift=0.23\textwidth]
\node[anchor=south west,inner sep=0](image) at (0,0){
\includegraphics[width=0.25 \textwidth]{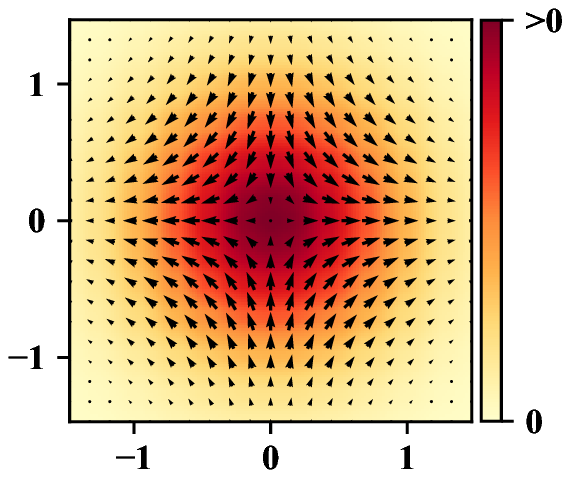}};
 \begin{scope}[x={(image.south east)},y={(image.north west)}]
 \node at (0.06,0.89) { $\mathbf{\left.{b}\right)}$};
 \node at (0.48,-0.02) {$\bm{x/l}$};
 \node at (0.04,0.45) {\begin{rotate}{90}$\bm{y/l}$\end{rotate}};
 \end{scope}
\end{scope}

\begin{scope}[xshift=0.00\textwidth,yshift=0.00\textwidth]
\node[anchor=south west,inner sep=0](image) at (0,0){
\includegraphics[width=0.25\textwidth]{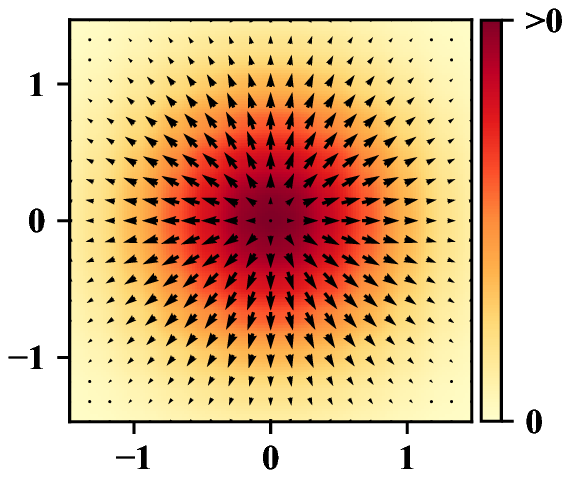}};
 \begin{scope}[x={(image.south east)},y={(image.north west)}]
 \node at (0.06,0.89) { $\mathbf{\left.{c}\right)}$};
 \node at (0.48,-0.02) {$\bm{x/l}$};
 \node at (0.04,0.45) {\begin{rotate}{90}$\bm{y/l}$\end{rotate}};
 \end{scope}
\end{scope}

\begin{scope}[xshift=0.25\textwidth,yshift=0.00\textwidth]
\node[anchor=south west,inner sep=0](image) at (0,0){
\includegraphics[width=0.25 \textwidth]{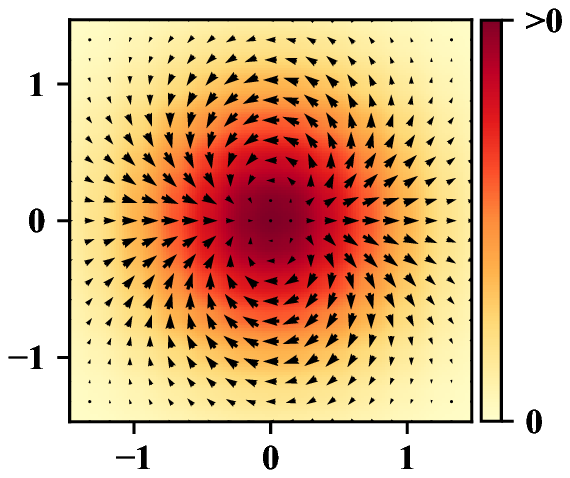}};
 \begin{scope}[x={(image.south east)},y={(image.north west)}]
 \node at (0.06,0.89) { $\mathbf{\left.{d}\right)}$};
 \node at (0.48,-0.02) {$\bm{x/l}$};
 \node at (0.04,0.45) {\begin{rotate}{90}$\bm{y/l}$\end{rotate}};
 \end{scope}
\end{scope}

\end{tikzpicture}
\par\end{centering}
\caption{Ground state in-plane spin textures $\boldsymbol{\sigma}_{\bot}$ (arrows)
presented on top of a density plot of $\varrho_{3}$. The results are
obtained for the non-trivial topological fields $\mathbf{B}_{n}^{\pm}$
and a symmetric harmonic confinement $\hbar\omega=5.0$ meV, and
$B_{0}=3.0$ T for \textbf{a)} $\mathbf{B}_{2}^{-}$ with winding number $q=-2$,
\textbf{b)} $\mathbf{B}_{1}^{-}$ with $q=-1$,
\textbf{c)} $\mathbf{B}_{1}^{+}$ with $q=1$, and \textbf{d)} $\mathbf{B}_{2}^{+}$ with $q=2$.}
\label{hs2}
\end{figure}

We exemplify the validity of Eq.~\eqref{eqPhaB02} for the following
magnetic fields which have a non-trivial topological pattern and
induce vortices with $q\in\mathbb{Z}$. These magnetic fields can, on
the one hand, arise due to magnetic textures in the plane which are
Hund-coupled with the spin of the electron. If, on the other hand,
$\mathbf{B}$ is an external magnetic field, then it must be
divergence-free, $\nabla\cdot\mathbf{B}=0$, as we are not interested
in analyzing the spin textures of magnetic monopoles. Besides,
$\mathbf{B}$ must be curl-free, $\nabla\times\mathbf{B}=0$, according
to the Maxwell equations since the model is taken to be
time-independent. The magnetic field
\begin{eqnarray}
\mathbf{B}_{n}^{-}=B_{0}\left(\frac{\rho}{l}\right)^{n}\left(\cos{(n\theta)},-\sin{(n\theta)},0\right)\label{eq:Bn0},
\label{eq_Bn}
\end{eqnarray}
fulfills the latter conditions and generates a vortex with a charge
$q=-n$. Here $n$ is a positive integer $n\in\mathbb{Z}^{+}$, and
$l=\sqrt{\hbar/(m_{e}\omega)}$ is the confinement length of the
potential $V_{0}=(1/2)m_{e}\omega^{2}\rho^{2}$. The magnetic field
\begin{eqnarray}
\mathbf{B}_{n}^{+}=B_{0}\left(\frac{\rho}{l}\right)^{n}\left(\cos{(n\theta)},\sin{(n\theta)},zf(x,y)\right),\,\,\,\label{eq:Bp}
\end{eqnarray}
on the other hand, has an in-plane vortex with a charge $q=n$ and
$n\in\mathbb{Z}^{+}$.  $f(x,y)$ is chosen such that the field is
divergence-free, where $x,y,z$ are Cartesian coordinates in a 3D
space. $\mathbf{B}_{n}^{+}$, however, can have a non-zero curl,
$\nabla\times\mathbf{B}_{n}^{+}\neq0$, which generates a
time-dependent electric field. To remedy this problem, we consider the
Hamiltonian in the limit $\hbar\omega_{c}/E\rightarrow0$ while
$(\Delta B_{0})/E$ is kept finite. In this limit, the induced
time-dependent electric field and the vector potential vanish and
leave a stationary system. The ground-state spin textures of an
electron exposed to $\mathbf{B}_{n}^{\pm}$ are shown in Fig.~\ref{hs2}
and are obtained numerically for $n=\pm1,\pm2$.

It is curious to note that the 2D magnetic field $\mathbf{B}_{n}^{-}$
can be curl and divergence free but $\mathbf{B}_{n}^{+}$ requires the
third field component to satisfy the constraints posed by the Maxwell
equations. This difference can be understood as follows:
$\mathbf{B}_{n}^{-}$ and the in-plane part of $\mathbf{B}_{n}^{+}$ can
be written in terms of complex variables $\varepsilon=x+iy$ and $\bar{\varepsilon}=x-iy$
as
\begin{eqnarray}
    \mathbf{B}_{n}^{-}=\frac{B_{0}}{l^{n}} \bar{\varepsilon}^{n}, \quad \mathbf{B}_{n\bot}^{+}=\frac{B_{0}}{l^{n}}\varepsilon^{n},
\end{eqnarray}
which shows that in this complex notation both fields are holomorphic \cite{Nakahara}
\begin{eqnarray}
    \frac{d}{d \varepsilon} \mathbf{B}_{n}^{-}=\frac{d}{d\bar{\varepsilon}} \mathbf{B}_{n\bot}^{+}=0.
\end{eqnarray}
On the other hand, if they are required to be simultaneously curl and divergence free, they have to also satisfy
\begin{eqnarray}
    \frac{d}{d \varepsilon} \mathbf{B}_{n}^{-}=\frac{d}{d\varepsilon} \mathbf{B}_{n\bot}^{+}=0.
\end{eqnarray}
This condition is trivially satisfied for $\mathbf{B}_{n}^{-}$ but it
forces the in-plane part of $\mathbf{B}_{n}^{+}$ to be identically
zero. Hence, the field $\mathbf{B}_{n}^{+}$ cannot just consist of an
$x$ and $y$ component; a non-zero $z$ component is needed.

\begin{figure}[tbh]
\begin{centering}
\begin{tikzpicture}

\begin{scope}[xshift=0.00\textwidth,yshift=0.00\textwidth]
\node[anchor=south west,inner sep=0](image) at (0,0){
\includegraphics[width=0.25\textwidth]{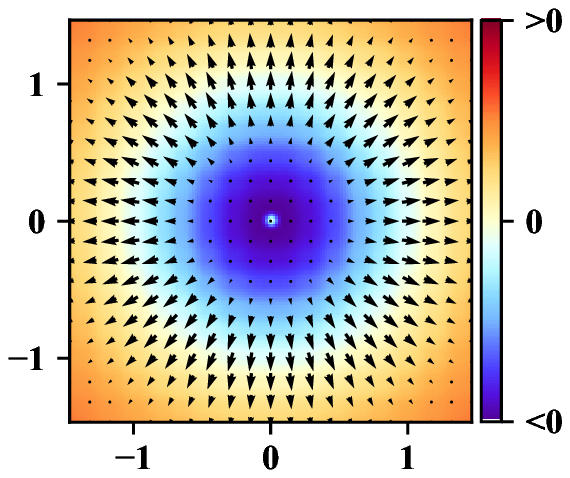}};
 \begin{scope}[x={(image.south east)},y={(image.north west)}]
 \node at (0.06,0.89) { $\mathbf{\left.{c}\right)}$};
 \node at (0.48,-0.02) {$\bm{x/l}$};
 \node at (0.04,0.45) {\begin{rotate}{90}$\bm{y/l}$\end{rotate}};
 \end{scope}
\end{scope}

\begin{scope}[xshift=0.25\textwidth,yshift=0.00\textwidth]
\node[anchor=south west,inner sep=0](image) at (0,0){
\includegraphics[width=0.25 \textwidth]{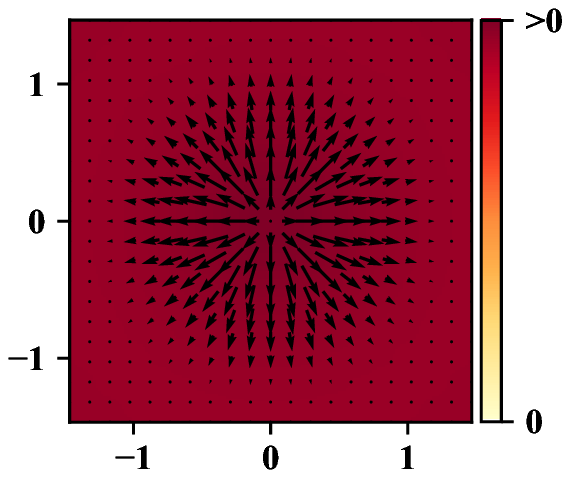}};
 \begin{scope}[x={(image.south east)},y={(image.north west)}]
 \node at (0.06,0.89) { $\mathbf{\left.{d}\right)}$};
 \node at (0.48,-0.02) {$\bm{x/l}$};
 \node at (0.04,0.45) {\begin{rotate}{90}$\bm{y/l}$\end{rotate}};
 \end{scope}
\end{scope}

\begin{scope}[xshift=0.00\textwidth,yshift=0.23\textwidth]
\node[anchor=south west,inner sep=0](image) at (0,0){
\includegraphics[width=0.25\textwidth]{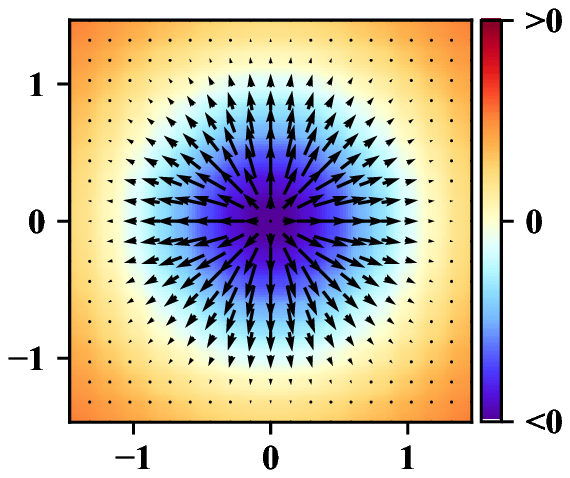}};
 \begin{scope}[x={(image.south east)},y={(image.north west)}]
 \node at (0.06,0.89) { $\mathbf{\left.{a}\right)}$};
 \node at (0.48,-0.02) {$\bm{x/l}$};
 \node at (0.04,0.45) {\begin{rotate}{90}$\bm{y/l}$\end{rotate}};
 \end{scope}
\end{scope}

\begin{scope}[xshift=0.25\textwidth,yshift=0.23\textwidth]
\node[anchor=south west,inner sep=0](image) at (0,0){
\includegraphics[width=0.25 \textwidth]{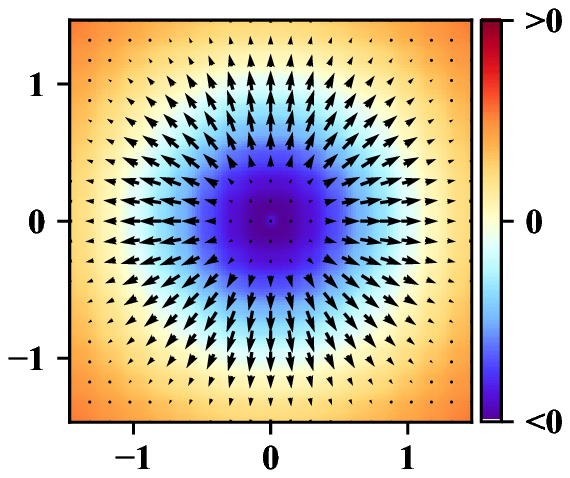}};
 \begin{scope}[x={(image.south east)},y={(image.north west)}]
 \node at (0.06,0.89) { $\mathbf{\left.{b}\right)}$};
 \node at (0.48,-0.02) {$\bm{x/l}$};
 \node at (0.04,0.45) {\begin{rotate}{90}$\bm{y/l}$\end{rotate}};
 \end{scope}
\end{scope}

\end{tikzpicture}
\par\end{centering}
\caption{In-plane spin textures $\boldsymbol\sigma_\bot$ presented on top of
the density plot of $\bar{\protect\varrho}_3 =\protect\varrho
_{3}/\protect\varrho _{0}$ for the Skyrmion field $\mathbf{B}^{s}$
with a symmetric harmonic confinement $\hbar \protect\omega =5.0$ meV,
$R=1.9l,$ $B_{0}=7.0$ T and $\protect\kappa =0.30$. Spin textures in
all the eigenstates have $q=1$, but the ground-state and the first and
second excited states have a Skyrmion number $Q=-1$ [see panels
\textbf{a)}, \textbf{b)}, and \textbf{c)}] while \textbf{d)} the third excited state has no Skyrmion.}
\label{fig_STSkyCont01}
\end{figure}

Next, we present the spin texture induced by a Skyrmion magnetic field. The magnetic field of a Skyrmion with $Q_{B}=-1$
has the form
\begin{eqnarray}
\mathbf{B}^{s}=B_{0}\left(\frac{2\kappa Rx}{\rho^{2}+\left(\kappa R\right){}^{2}},\frac{2\kappa Ry}{\rho^{2}+\left(\kappa R\right){}^{2}},\frac{\rho^{2}-\left(\kappa R\right){}^{2}}{\rho^{2}+\left(\kappa R\right){}^{2}}\right), \,\,\,\,
\label{eq_Skyrm}
\end{eqnarray}
where $\kappa$ is a dimensionless parameter and $R$ is a constant with a dimension of length.
The spin textures of a few low-energy eigenstates in the presence of
the field $\mathbf{B}^s$ are presented in
Fig.~\ref{fig_STSkyCont01}. Although the in-plane winding number is
$q=1$ for all the states, not all of them have the same Skyrmion
charges: the third excited state has no Skyrmion while the other
states have a Skyrmion with $Q=-1$.

\section{Magnetic fields with a trivial topological pattern}
\label{SecMFWTTP}

In this section, we explore magnetic fields which are topologically
trivial in the sense that $q_B=Q_B=0$ but induce non-trivial spin
textures $q\neq 0$ or $Q\neq 0$.

\subsection{Continuous Magnetic Fields}
\label{SubSecMFWTTP01}

First, we focus on a class of systems experiencing a magnetic field
$\mathbf{B}^{p}=(B_{1},0,B_{3})$ with $Q_{B}=q_{B}=0$ which can be
thought of as being the projection of a topologically non-trivial
field $\mathbf{B}=(B_{1},B_{2},B_{3})$ onto the $x-z$ plane. Our
analysis of the continuity equations for this case can qualitatively
explain the emergent spin textures. Furthermore, it is in agreement
with perturbation theory and exact numerical results, as presented
below.
We study the continuity equations in the asymptotic limit $\rho\gg
L_{x,y}$. The in-plane vector potential is fixed to
$\mathbf{A}=(B_{0}/2)(-y,x,0)$. The spin continuity equations are then approximated
again by Eq.~(\ref{eqSCE02}). In WKB spirit, we expand the spin fields
in the smallest energy scale of the theory $\bar{\omega}_{c}=\hbar \omega_{c}/E$, see
Eq.~\eqref{spinexpand}, and now keep terms up to linear order in
$\bar{\omega}_{c}$. In zeroth order we find ($\tilde{B}_{j}=\Delta {B}_{j}/E$)
\begin{eqnarray}
\label{zeroth_order_WKB}
\varrho_{3}^{(0)}\tilde{B}_{1}=\varrho_{1}^{(0)} \tilde{B}_{3},\text{ and }\varrho_{2}^{(0)}=0,
\end{eqnarray}
which shows that in this semiclassical approximation the spin
densities, as expected, follow the magnetic field. Since the magnetic
field considered is topologically trivial, the spin fields at this
order are topologically trivial as well.

At the first order in $\bar{\omega}_{c}$ we find
\begin{equation}
\label{first_order_WKB}
\frac{\varrho_1^{(1)}}{\varrho_3^{(1)}}=\frac{\tilde{B}_1}{\tilde{B}_3},
\quad \varrho_2^{(1)}=-\frac{1}{\tilde{B}_{3}}\partial_\theta \varrho_1^{(0)}
=\frac{1}{ \tilde{B}_1}\partial_\theta \varrho_3^{(0)} \, .
\end{equation}
Thus the fields $\varrho_{1,3}$ still follow the magnetic field. The
field $\varrho_2$, however, is no longer zero. To solve these
equations, we can set $\varrho_{1,3}^{(0)}=\varrho^{(0)}
\tilde{B}_{1,3}$ with an unknown function $\varrho^{(0)} \neq 0$.
This solves the zeroth order equation \eqref{zeroth_order_WKB}. Using
this lowest order result in the second relation of
Eq. \eqref{first_order_WKB} and assuming that $\tilde{B}_3$ is
independent of the angle $\theta$ we obtain the following differential
equation
\begin{equation}
\label{WKB2}
\frac{\tilde{B}_1\partial_\theta \tilde{B}_1}{\tilde{B}_1^2+\tilde{B}_3^2} = -\frac{\partial_\theta \varrho^{(0)}}{\varrho^{(0)}}
\end{equation}
which gives $\varrho^{(0)}=\pm F(\rho) (\tilde{B}_1^2+\tilde{B}_3^2)^{-1/2}$ where $F(\rho)$ is the radial part of $\varrho^{(0)}$. Using Eq.~\eqref{first_order_WKB} we can now also
determine $\varrho_2^{(1)}$. To lowest non-vanishing order we thus find
\begin{eqnarray}
\label{WKB3}
\varrho_3 &=& \pm F(\rho) \frac{\tilde{B}_{3}}{\sqrt{\tilde{B}_1^2+\tilde{B}_3^2}},\quad \varrho_1=\pm F(\rho)\frac{\tilde{B}_{1}}{\sqrt{\tilde{B}_1^2+\tilde{B}_3^2}},\nonumber \\
\varrho_2 &=& \mp \bar{\omega}_c F(\rho) \frac{\tilde{B}_{3}\partial_\theta \tilde{B}_1}{(\tilde{B}_1^2+\tilde{B}_3^2)^{3/2}} \,.
\end{eqnarray}

Eq.~(\ref{WKB3}) shows that quantum corrections lead to a non-zero
component of the spin-field in $y$-direction despite the fact that the
$y$-component of the magnetic field $\mathbf{B}^{p}$ is
zero. Therefore a topologically trivial field such as, for example,
\begin{eqnarray}
\mathbf{B}_{n}^{p}=B_{0}(f(\rho)\cos{(n\theta)},0,1),
\end{eqnarray}
does generate a vortex in the spin densities (with $|q|=|n|$ in this
case).
 To understand the sign of the winding number, we can treat the
 $x$-component of $\mathbf{B}_{n}^{p}$ with $f=(b\rho)^{|n|}$ (in
 which $b$ has a dimension of inverse length) as a perturbation in the
 Hamiltonian with a symmetric harmonic trap
 $V_{0}=\frac{1}{2}m_{e}\omega^{2}\rho^{2}$. In the symmetric gauge,
 the spin fields are then given by
\begin{equation}
\varrho_{1}=F_{1}(\rho)\cos{(n\theta)},\quad\varrho_{2}=\sigma F_{2}(\rho)\sin{(n\theta)},
\end{equation}
for an eigenfunction with spin quantum number $\sigma=\pm1$ (details
for the case $n=1$ are given in App. \ref{AppPer}). The winding number
of the induced vortex is $q=\sigma n\,\text{sgn}(F_{1}F_{2})$ for
$F_{1,2}\neq0$, and can take on $+n$ and $-n$ in different
eigenstates. This feature is contrary to the semiclassical cases
discussed in the previous section, in which the electron experiences a
magnetic field with a non-trivial pattern and the spin textures in {\it
all the eigenfunctions} have an identical topology. If the vector
potential is switched off (for instance by considering the regime
$\hbar
\omega_{c}/(\Delta B_{0}) \rightarrow0 $) in the presence of
$\mathbf{B}^{p}$, the unperturbed eigenfunctions can be written in
terms of Hermite polynomials which are real functions, and the spin
texture is thus topological trivial. We note that this is again
consistent with the WKB approximation, Eq.~\eqref{WKB3}.

\begin{figure}[!htb]
\begin{centering}
\begin{tikzpicture}

\begin{scope}[xshift=0.00\textwidth,yshift=0.00\textwidth]
\node[anchor=south west,inner sep=0](image) at (0,0){
\includegraphics[width=0.25\textwidth]{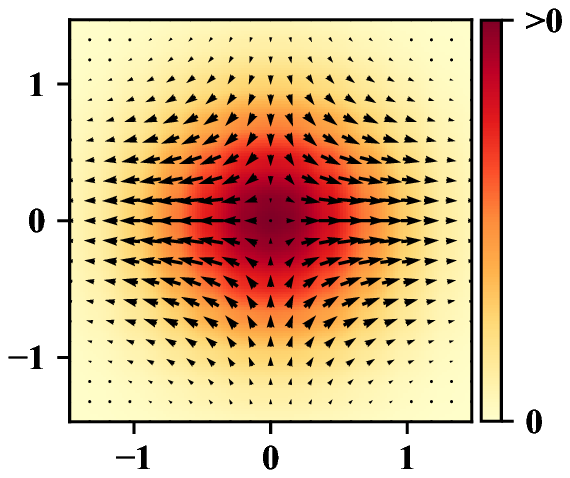}};
 \begin{scope}[x={(image.south east)},y={(image.north west)}]
 \node at (0.06,0.89) { $\mathbf{\left.{a}\right)}$};
 \node at (0.48,-0.02) {$\bm{x/l}$};
 \node at (0.04,0.45) {\begin{rotate}{90}$\bm{y/l}$\end{rotate}};
 \end{scope}
\end{scope}

\begin{scope}[xshift=0.25\textwidth,yshift=0.00\textwidth]
\node[anchor=south west,inner sep=0](image) at (0,0){
\includegraphics[width=0.25 \textwidth]{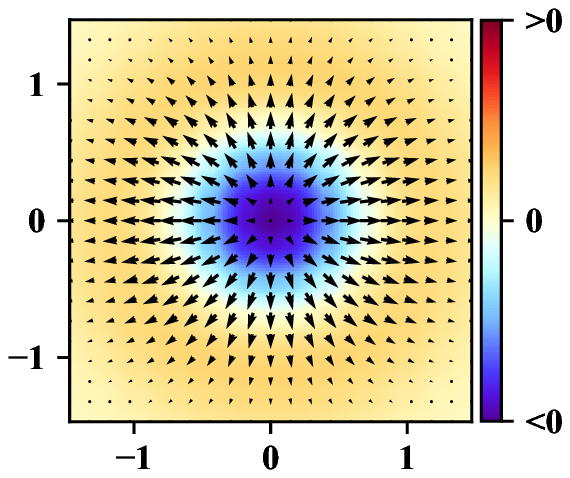}};
 \begin{scope}[x={(image.south east)},y={(image.north west)}]
 \node at (0.06,0.89) { $\mathbf{\left.{b}\right)}$};
 \node at (0.48,-0.02) {$\bm{x/l}$};
 \node at (0.04,0.45) {\begin{rotate}{90}$\bm{y/l}$\end{rotate}};
 \end{scope}
\end{scope}

\end{tikzpicture}
\par\end{centering}
\caption{In-plane spin textures $\boldsymbol{\sigma}_{\bot}$
presented on top of the density plot of $\bar{\varrho}_{3}$ for a
model with symmetric harmonic confinement and a magnetic field
$\mathbf{B}_{1}^{p}$ where $B{_{0}}=3.0$ T and $\hbar\omega=5.0$
meV. \textbf{a)} The spin texture is a vortex with
$q=-1$ in the ground state, and \textbf{b)} a Skyrmion with Q=-1 and vorticity $q=1$ in
the first excited state.}
\label{fig_STBP01}
\end{figure}

\begin{figure}[!htb]
\begin{centering}
\begin{tikzpicture}

\begin{scope}[xshift=0.00\textwidth,yshift=0.00\textwidth]
\node[anchor=south west,inner sep=0](image) at (0,0){
\includegraphics[width=0.25\textwidth]{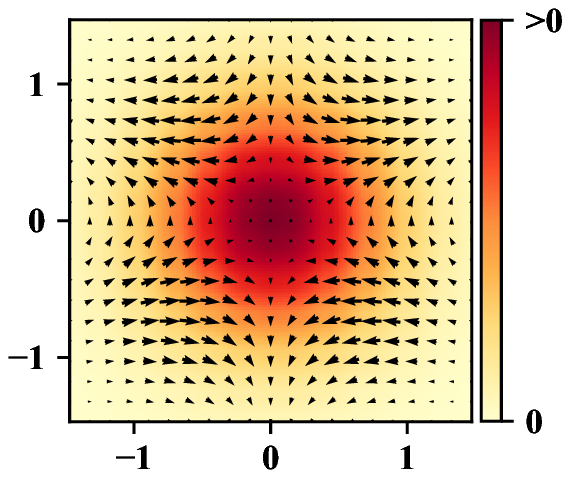}};
 \begin{scope}[x={(image.south east)},y={(image.north west)}]
 \node at (0.06,0.89) { $\mathbf{\left.{a}\right)}$};
 \node at (0.48,-0.02) {$\bm{x/l}$};
 \node at (0.04,0.45) {\begin{rotate}{90}$\bm{y/l}$\end{rotate}};
 \end{scope}
\end{scope}

\begin{scope}[xshift=0.25\textwidth,yshift=0.00\textwidth]
\node[anchor=south west,inner sep=0](image) at (0,0){
\includegraphics[width=0.25 \textwidth]{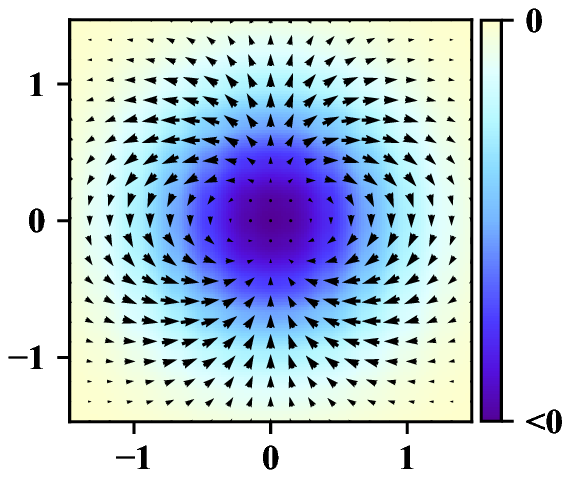}};
 \begin{scope}[x={(image.south east)},y={(image.north west)}]
 \node at (0.06,0.89) { $\mathbf{\left.{b}\right)}$};
 \node at (0.48,-0.02) {$\bm{x/l}$};
 \node at (0.04,0.45) {\begin{rotate}{90}$\bm{y/l}$\end{rotate}};
 \end{scope}
\end{scope}

\end{tikzpicture}
\par\end{centering}
\caption{Same as Fig.~\ref{fig_STBP01} but for the field $\mathbf{B}_{2}^{p}$
where $B{_{0}}=2.9$ T and $\hbar\omega=5.0$ meV. The spin textures are
now vortices with \textbf{a)} $q=-2$ in the ground state, and
\textbf{b)} $q=2$ in the first excited state.}
\label{fig_STBP02}
\end{figure}

To exemplify the generation of topological spin textures in
topologically trivial magnetic fields of the type $\mathbf{B}^{p}_n$, we
numerically calculate the eigenfunctions for a symmetric harmonic trap
$V_{0}$ with a length scale $l=\sqrt{\hbar/(m_{e}\omega)}$. In Fig.~\ref{fig_STBP01}, the spin textures of the ground state and the
first excited state induced by the magnetic field
\begin{eqnarray}
\mathbf{B}_{1}^{p}=B_{0}(\frac{x}{l},0,1),
\end{eqnarray}
are shown which are vortices with $q=-1$ ($q=1$) in the ground state
(the first excited state), respectively. This field is curl-free and
can be made divergence-free by adding a field $B_{0}bz\,\hat{z}$
to $\mathbf{B}_{1}^p$, and by noting that the plane in which the
electron moves is at $z=0$. It is interesting to note that the spin
texture of the first excited state under $\mathbf{B}_{1}^p$ is a
Skyrmion with $Q=-1$. As a second example, we consider the spin textures induced by
\begin{eqnarray}
\mathbf{B}_{2}^{p}=B_{0}(\frac{xy}{l^{2}},0,1) \, .
\end{eqnarray}
As shown in Fig.~\ref{fig_STBP02} we now obtain vortices with $q=-2$
($q=2$) in the ground state (the first excited state).

\subsection{Discontinuous Magnetic Fields}
\label{SubSecMFWTTP02}

The second type of magnetic fields discussed in this section are 
discontinuous
functions of $\mathbf{r}$. We are not aiming at presenting a general
analysis of all possible kinds of discontinuous fields here but will
rather limit ourselves to a couple of helpful examples.

The first example we want to discuss, is a field $\mathbf{B}_{n}^{d}$
with components
\begin{eqnarray}
\mathbf{B}_{n}^{d}\cdot\hat{x}=\alpha B_{0}\left(\Theta(\cos(n\theta))-\frac{1}{2}\right),\\
\mathbf{B}_{n}^{d}\cdot\hat{y}=\beta B_{0}\left(\Theta(\sin(n\theta))-\frac{1}{2}\right),
\end{eqnarray}
where $\alpha,\beta$ are dimensionless constants and $\Theta(x)$ is
the Heaviside step function. Such fields can be thought of as arising
from magnetic domains inside the plane to which the electron is
confined \cite{Nogaret,inho01}. The discontinuity of the Heaviside
step function $\Theta(x)$ can be smoothed out by using the inverse
tangent function $\sim\tan^{-1}(\lambda x)$ instead because the jump
at $x=0$ takes place within a distance which is much smaller than the
relevant length scale for the electron, $\lambda L_{x,y}\gg 1$. Using
the inverse tangent function instead of the Heaviside step function
makes it possible to use Eq.~(\ref{eqPhaB03}) to show that
$\mathbf{B}_{n}^{d}$ induces an in-plane vortex with winding number
$q=n$. In the limit $\lambda L_{x,y}\gg1$, however, the vector
potential $\mathbf{A}$ is required in order to soften the
discontinuity. Otherwise, the spin texture is trivial since the spin
fields cannot be discontinuous.

\begin{figure}[!htb]
\begin{centering}
\begin{tikzpicture}

\begin{scope}[xshift=0.00\textwidth,yshift=0.00\textwidth]
\node[anchor=south west,inner sep=0](image) at (0,0){
\includegraphics[width=0.25\textwidth]{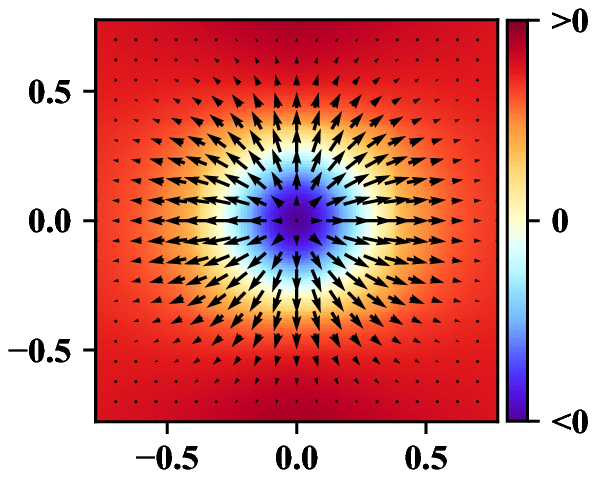}};
 \begin{scope}[x={(image.south east)},y={(image.north west)}]
 \node at (0.06,0.89) { $\mathbf{\left.{a}\right)}$};
 \node at (0.48,-0.02) {$\bm{x/l}$};
 \node at (0.04,0.45) {\begin{rotate}{90}$\bm{y/l}$\end{rotate}};
 \end{scope}
\end{scope}

\begin{scope}[xshift=0.25\textwidth,yshift=0.00\textwidth]
\node[anchor=south west,inner sep=0](image) at (0,0){
\includegraphics[width=0.25 \textwidth]{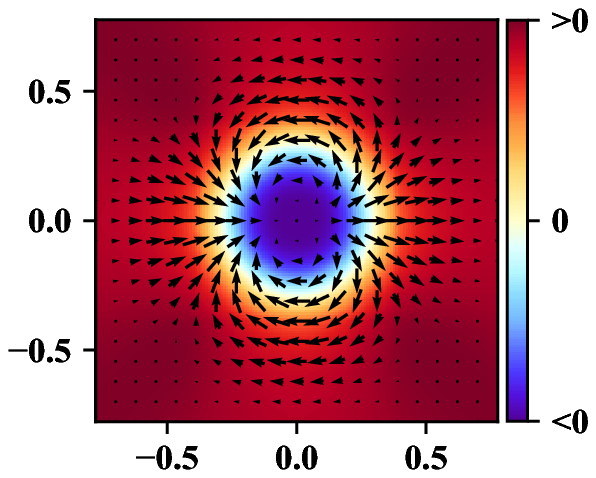}};
 \begin{scope}[x={(image.south east)},y={(image.north west)}]
 \node at (0.06,0.89) { $\mathbf{\left.{b}\right)}$};
 \node at (0.48,-0.02) {$\bm{x/l}$};
 \node at (0.04,0.45) {\begin{rotate}{90}$\bm{y/l}$\end{rotate}};
 \end{scope}
\end{scope}

\end{tikzpicture}
\par\end{centering}
\caption{Ground-state in-plane spin textures $\boldsymbol{\sigma}_{\bot}$
presented on top of the spin density $\varrho_{3}$ for the magnetic
field $\mathbf{B}_{n}^{d}$ with $L=36$ nm, $\kappa=0.32$, $\alpha=1$,
$\beta=0$, and $\gamma=2$. Non-zero Skyrmion and winding numbers are
found, for example, for \textbf{a)} $B{_{0}}=5.0$ T and $n=1$, with charges
$Q=-1$ and $q=1$, and \textbf{b)} for $B{_{0}}=7.7$ T, $n=2$ where $Q=-2$ and
$q=2$.}
\label{fig_SPSky02}
\end{figure}

It is also possible to turn the spin texture of the electron into
a Skyrmion with $|Q|=|n|$ if the $z$-component of $\mathbf{B}_{n}^{d}$
is tailored to
\begin{eqnarray}
\mathbf{B}_{n}^{d}\cdot\hat{z}=\gamma B_{0}\left(\Theta\left({\rho}/{L}-\kappa\right)-\frac{1}{2}\right),
\end{eqnarray}
where $L$ is a length scale, and $\kappa$ is a dimensionless constant.
Furthermore, it is interesting to note that even after switching off
the $y$-component of $\textbf{B}^{d}$ (i.e. $\beta=0$), the
topological spin textures persist as shown in Fig.~\ref{fig_SPSky02}.

\begin{figure}[!htb]
\begin{centering}
\begin{tikzpicture}

\begin{scope}[xshift=0.00\textwidth,yshift=0.00\textwidth]
\node[anchor=south west,inner sep=0](image) at (0,0){
\includegraphics[width=0.25\textwidth]{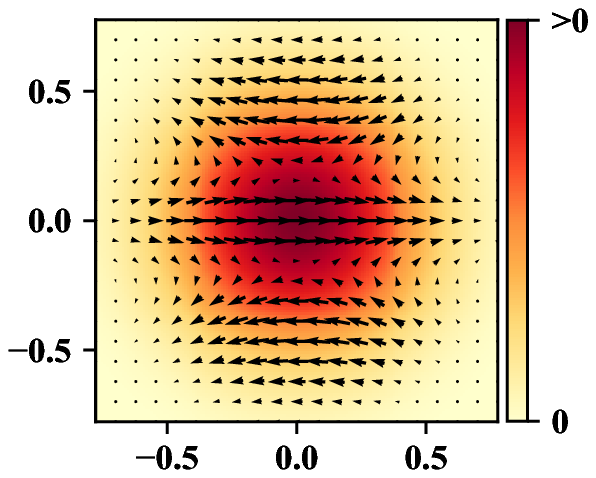}};
 \begin{scope}[x={(image.south east)},y={(image.north west)}]
 \node at (0.06,0.89) { $\mathbf{\left.{a}\right)}$};
 \node at (0.48,-0.02) {$\bm{x/R}$};
 \node at (0.04,0.45) {\begin{rotate}{90}$\bm{y/R}$\end{rotate}};
 \end{scope}
\end{scope}

\begin{scope}[xshift=0.25\textwidth,yshift=0.00\textwidth]
\node[anchor=south west,inner sep=0](image) at (0,0){
\includegraphics[width=0.25 \textwidth]{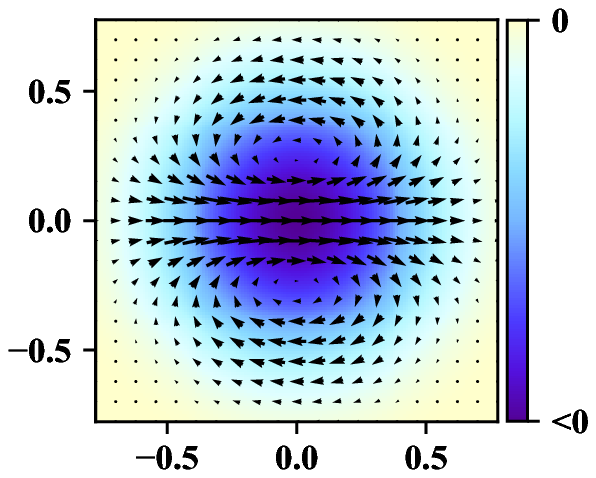}};
 \begin{scope}[x={(image.south east)},y={(image.north west)}]
 \node at (0.06,0.89) { $\mathbf{\left.{b}\right)}$};
 \node at (0.48,-0.02) {$\bm{x/R}$};
 \node at (0.04,0.45) {\begin{rotate}{90}$\bm{y/R}$\end{rotate}};
 \end{scope}
\end{scope}

\end{tikzpicture}
\par\end{centering}
\caption{In-plane spin textures $\boldsymbol{\sigma}_{\bot}$
presented on top of the spin density $\varrho_{3}$ for the magnetic
field $\mathbf{B}_{||}$ with $R=36$ nm as the radius of the potential
well, and $B{_{0}}=2.9$ T. The parameters of the magnetic domain are
$\mu=2$, $\nu=1$ and $\kappa=0.15$. The spin texture is a vortex with
\textbf{a)} $q=-2$ in the ground state, and \textbf{b)} with $q=2$ in the
first excited state.}
\label{fig_SPDDW02}
\end{figure}

As a second example, we present the spin textures of an electron in a
hard-wall potential
\begin{eqnarray}
V_{hw}=\begin{cases}
0 & \rho<R,\\
\infty & \rho\geqslant R,
\end{cases}
\end{eqnarray}
experiencing the
discontinuous in-plane magnetic field
\begin{eqnarray}
\mathbf{B}_{||}=\begin{cases}
{B}_{0}(-\nu,0,0) & y\geqslant\kappa R,\\
{B}_{0}(\mu,0,0) & -\kappa R<y<\kappa R,\\
{B}_{0}(-\nu,0,0) & y\leqslant-\kappa R,
\end{cases}
\end{eqnarray}
where the domain walls are now parallel to each other rather than
radially arranged as in the first example. The parameters chosen to
illustrate the spin textures are $\mu=2$, $\nu=1$, and
$\kappa=0.15$. We include, furthermore, a uniform magnetic field
$\mathbf{B}_{0}=B_{0}\hat{z}$ perpendicular to the plane which is
generated by the vector potential $\mathbf{A}=(B_{0}/2)(-y,x,0)$. The
spin textures of the ground state and the first excited state for this
case are depicted in Fig.~\ref{fig_SPDDW02}. A vortex with $q=-2$ is
present in the ground state, and a vortex with $q=2$ in the first
excited state. We note that in this setup, the energies confined to
each of the domains need to be of the same order of magnitude to
favour the formation of vortices.

\begin{figure}[!htb]
\begin{centering}
\begin{tikzpicture}

\begin{scope}[xshift=0.00\textwidth,yshift=0.00\textwidth]
\node[anchor=south west,inner sep=0](image) at (0,0){
\includegraphics[width=0.25\textwidth]{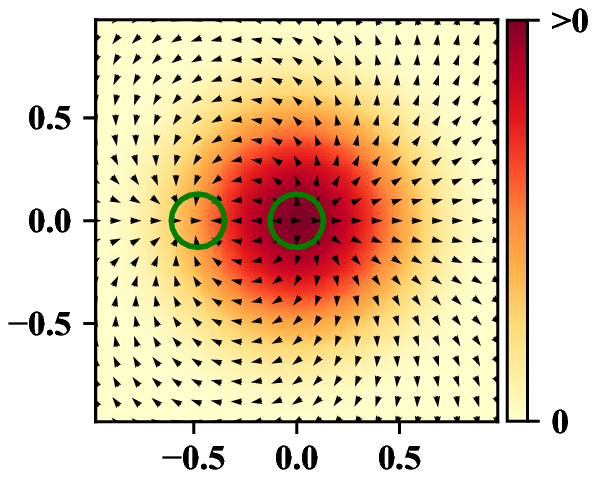}};
 \begin{scope}[x={(image.south east)},y={(image.north west)}]
 \node at (0.06,0.89) { $\mathbf{\left.{a}\right)}$};
 \node at (0.48,-0.02) {$\bm{x/R}$};
 \node at (0.04,0.45) {\begin{rotate}{90}$\bm{y/R}$\end{rotate}};
 \end{scope}
\end{scope}

\begin{scope}[xshift=0.25\textwidth,yshift=0.00\textwidth]
\node[anchor=south west,inner sep=0](image) at (0,0){
\includegraphics[width=0.25 \textwidth]{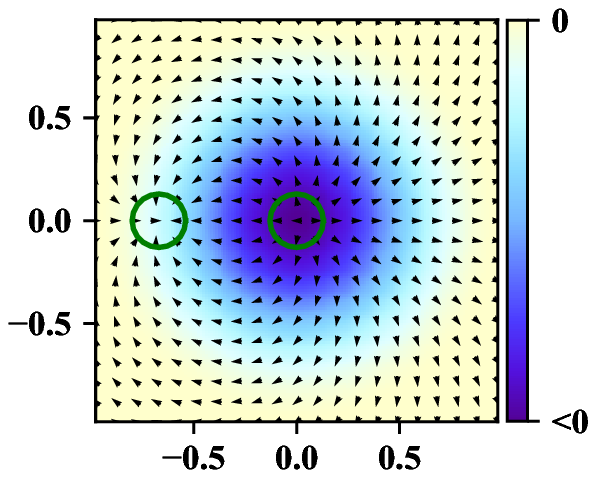}};
 \begin{scope}[x={(image.south east)},y={(image.north west)}]
 \node at (0.06,0.89) { $\mathbf{\left.{b}\right)}$};
 \node at (0.48,-0.02) {$\bm{x/R}$};
 \node at (0.04,0.45) {\begin{rotate}{90}$\bm{y/R}$\end{rotate}};
 \end{scope}
\end{scope}

\end{tikzpicture}
\par\end{centering}
\caption{In-plane spin textures $\boldsymbol{\sigma}_{\bot}$
presented on top of the spin density $\varrho_{3}$ for the magnetic
field $\mathbf{B}_{12}$ with $R=36$ nm as the radius of the potential
well, $B{_{0}}=1.0$ T, $b=0.25$ R, and $\alpha=\beta=0.2$. The net
winding number is $q=2$ in \textbf{a)} the ground state, as well as
\textbf{b)} in the first excited state. The vortices, however, now have
two singularities. Their positions are indicated by circles.}
\label{fig_SPMixed01}
\end{figure}

\section{Magnetic fields with piecewise-defined winding numbers}
\label{SecNoGolTo}
Finally, we will also briefly explore the spin textures in the
presence of a magnetic field which in different domains of the plane
has a different topological pattern and thus cannot be fully described
by a single and unique winding number. This is possible because our theory
is not scale free and at different length scales, different fields can
be dominant.

We assume the in-plane magnetic field $\mathbf{B}_{\bot}$ in the Hamiltonian
is a vector sum of fields and each field has a definite winding number.
Equivalently, in terms of $\varepsilon$ and $\bar{\varepsilon}$, we write
\begin{eqnarray}
	\mathbf{B}_{\bot}
	=
	B_{0}
	\sum_{i=1 }
	\left(\frac{\varepsilon}{b^{+}_{i}}\right)^{n^{+}_{i}}
	+
	B_{0}
	\sum_{i=1  }
	\alpha^{-}_{i}
	\left(\frac{\bar{\varepsilon}}{b^{-}_{i}}\right)^{n^{-}_{i}},
\end{eqnarray}
and a finite set of positive integers $\{n^{\pm}_1,n^{\pm}_2,n^{\pm}_3,\dots  \}$ defines
the magnetic fields and $b^{\pm}_{i}$ are constants with a dimension of length.
The sum of complex variables can be solved locally and be written
as a single-term complex variable $\mathbf{B}_{\bot}=B_{0}\chi  e^{iM\theta}$
where $\chi  \equiv \chi (\mathbf{r})$ and $M$ is an integer which is defined locally
$M \equiv M(\mathbf{r})$. Eq. (\ref{eqPhaB02}) can be used to find the winding numbers at different domains.
The net winding number is given by the largest $n^{\pm}_{i}$. The singularities of the magnetic fields can be found by solving the equation $\mathbf{B}_{\bot}=0$. However, the singularities of the spin texture can be redistributed since the magnetic fields are renormalized by the terms in the Hamiltonian, for instance the vector potential.
As an example, we want to focus on a magnetic field which is a linear
combination of two fields,
$\mathbf{B}_{12}=\alpha \mathbf{B}_{1}+\beta \mathbf{B}_{2}$
($\alpha$ and $\beta$ are dimensionless constants), with different winding numbers $q$ and
\begin{eqnarray}
\mathbf{B}_{n}=B_{0}\left(\frac{\rho}{b}\right)^{n}\left(\cos{(n\theta)},\sin{(n\theta)},0\right).
\label{eq:Bn}
\end{eqnarray}
The normalized field
$\bar{\mathbf{B}}_{12}=\mathbf{B}_{12}/|\mathbf{B}_{12}|$ gives
$q^{<}=1$ in the region $\rho<b\alpha/\beta$ and $q^{>}=2$ for
$\rho>b\alpha/\beta$, but becomes singular at $\rho=b\alpha/\beta$. An
estimate for the position of the singularities of the spin texture can
be obtained by solving the equation $\mathbf{B}_{12}= 0$ but this
position is, in general, further renormalized depending on the
eigenstate of the electron.
This is demonstrated in the numerical results shown in
Fig.~\ref{fig_SPMixed01}. Here the position of the second singularity
does indeed change slightly for different eigenstates. We conclude
that the simultaneous presence of magnetic fields with different
topologies does not necessarily lead to a trivial spin texture, but
can rather split the spin texture into different domains. The
topological index of each domain is then associated with the one of
the dominant field in this domain.

\section{conclusion}
\label{Concl}

Our study shows that topologically non-trivial spin textures of a
single confined electron can emerge in the presence of inhomogeneous
magnetic fields. This includes in-plane vortices and three-dimensional
Skyrmion configurations. The fields can either be applied external
fields or effective fields which arise due to a magnetic background in
the plane the electron is confined to. Most remarkably, we find that
topological spin fields can form even if the magnetic field itself is
topologically trivial. This includes the experimentally relevant case
of magnetic domain walls.

To understand the spin textures analytically, we started from the
continuity equations for the spin fields. At distances large compared
to the length scale of the confining potential, the kinetic energy
term in the Hamiltonian becomes the smallest energy scale and the
continuity equation can be analyzed in the spirit of a WKB
approximation. In lowest order---which corresponds to a semiclassical
approximation---one finds, as expected, that the spin fields simply
follow the magnetic field. A magnetic field which itself has a
topologically non-trivial in-plane winding number thus always leads to
in-plane spin textures with the same topological index as the field
in all of the electronic eigenstates of the system. Interestingly, the
same does not apply to the Skyrmion number: Eigenstates of a system in
a magnetic field with a non-zero Skyrmion number do not necessarily
share this Skyrmion number. Here our WKB approximation turns out to be
insufficient because the Skyrmion number is determined by the
three-dimensional spin texture across the entire dot. The knowledge of
the spin arrangement for distances large compared to the confinement
length is not sufficient.

Even more surprisingly, we have demonstrated that quantum
corrections---obtained in higher orders of the WKB approximation---can
induce topologically non-trivial spin textures for certain
topologically trivial magnetic fields. In this case the topological
index is, however, not fixed and can vary between different
eigenfunctions of the system. The WKB analysis is supported by
perturbation theory and exact numerical results. We also examined the
case where the magnetic field has a piecewise-defined topological
pattern and have argued that the spin texture then also acquires a
piece-wise defined topological charge.

Our study is restricted to stationary systems and temperatures
$k_{B}T\ll\hbar\omega$ where thermal fluctuations can be
neglected. Topological pattern in the spin density of an electron are,
in principle, detectable in experiment. A first step in this direction
has been the recent detailed mapping of the charge density of a single
electron in a quantum dot \cite{PRL122}. Imprinting topological spin
textures in quantum dots using either external magnetic fields or by
tayloring domain walls in the plane the electron is confined to, might
be an interesting avenue to explore in the search for topological
spintronics applications.

\section{Acknowledgement}
W.L. acknowledges support by the NSF-China under Grant
No. 11804396. A.N. acknowledges partial financial support by the
Canada Research Chairs program through Tapash Chakraborty. A.N. and
J.S. acknowledge support by the Natural Sciences and Engineering
Research Council (NSERC) through the Discovery Grants program.

\appendix
\label{AppCurrDenEX}

\section{Topological densities in terms of current densities}

In this appendix, we derive identities which attribute the topological
charges to current densities by representing the current densities
in terms of the generic eigenfunction \eqref{eq_GEF}.

\subsection{Winding number in terms of current densities}

\label{AppWNCD} First, we define the canonical current as
\begin{eqnarray}
 &  & \mathbf{J}_{\mu}=\frac{\hbar}{2m_{e}}\tilde{\mathbf{J}}_{\mu}-\frac{e}{m_{e}}\mathbf{A}\varrho_{\mu},
 \label{eq_NewCurDen}
\end{eqnarray}
which have the following forms in term of the the generic eigenfunction
given in eq.~\eqref{eq_GEF}
\begin{eqnarray}
 &  & \tilde{\mathbf{J}}_{0}=\varrho_{0}\nabla\left(S_{1}+S_{2}\right)-\varrho_{3}\nabla\left(S_{2}-S_{1}\right),\\
 &  & \tilde{\mathbf{J}}_{1}=\varrho_{1}\nabla\left(S_{1}+S_{2}\right)-\varrho_{2}\nabla\ln{\left(\frac{\psi_{1}}{\psi_{2}}\right)},\\
 &  & \tilde{\mathbf{J}}_{2}=\varrho_{2}\nabla\left(S_{1}+S_{2}\right)+\varrho_{1}\nabla\ln{\left(\frac{\psi_{1}}{\psi_{2}}\right)},\\
 &  & \tilde{\mathbf{J}}_{3}=\varrho_{3}\nabla\left(S_{1}+S_{2}\right)-\varrho_{0}\nabla\left(S_{2}-S_{1}\right).
\end{eqnarray}
The identity $\mathbf{J}_{\nu}\eta^{\nu\mu}\varrho_{\mu}=0$ can be
immediately established
(summation on repeated indices and the Minkowski metric $\eta^{\nu\mu}$
with the signature $(+,-,-,-)$). Gradients of the fields have the following
forms
\begin{eqnarray}
 &  & \nabla\varrho_{0}=\varrho_{0}\nabla\ln{\left(\psi_{1}\psi_{2}\right)}+\varrho_{3}\nabla\ln{\left(\frac{\psi_{1}}{\psi_{2}}\right)},\\
 &  & \nabla\varrho_{1}=\varrho_{1}\nabla\ln{\left(\psi_{1}\psi_{2}\right)}-\varrho_{2}\nabla\left(S_{2}-S_{1}\right),\\
 &  & \nabla\varrho_{2}=\varrho_{2}\nabla\ln{\left(\psi_{1}\psi_{2}\right)}+\varrho_{1}\nabla\left(S_{2}-S_{1}\right),\\
 &  & \nabla\varrho_{3}=\varrho_{3}\nabla\ln{\left(\psi_{1}\psi_{2}\right)}+\varrho_{0}\nabla\ln{\left(\frac{\psi_{1}}{\psi_{2}}\right)}.
\end{eqnarray}
These equations can be used to derive the gradient of the phase difference
\begin{eqnarray}
\nabla\left(S_{2}-S_{1}\right) & = & \frac{\varrho_{1}\nabla\varrho_{2}-\varrho_{2}\nabla\varrho_{1}}{\varrho_{1}^{2}+\varrho_{2}^{2}}\\
 & = & \frac{2m_e}{\hbar}\frac{\varrho_{3}\mathbf{J}_{0}-\varrho_{0}\mathbf{J}_{3}}{\varrho_{0}^{2}-\varrho_{3}^{2}},
\end{eqnarray}
where in the second line, the identity $\varrho_{\nu}\eta^{\nu\mu}\varrho_{\mu}=0$ is also used.

\subsection{Skyrmion density in terms of current densities}

\label{AppTDCD} We use the generic eigenfunction in eq.~\eqref{eq_GEF}
(though in a different gauge obtained by changing the gauge of the
vector potential) in order to write down the topological density in
terms of the spin currents. We also use the $SU(2)$ parametrization
$(\vartheta,\varphi)$ to write the eigenfunction
\begin{eqnarray}
\tilde{\Psi}(\mathbf{r})=\begin{pmatrix}\psi_{1}\\
e^{iS}\psi_{2}
\end{pmatrix}=\sqrt{\varrho_{0}}\begin{pmatrix}\cos(\vartheta/2)\\
e^{i\varphi}\sin(\vartheta/2)
\end{pmatrix},
\label{eq_FigFNG}
\end{eqnarray}
where $S=S_{2}-S_{1}$ and the vector potential needs to be replaced
by $\mathbf{A}^{\prime}=\mathbf{A}-(\hbar/e)\nabla S_{1}$. The spin
fields become
\begin{eqnarray}
\varrho_{1} & = & \varrho_{0}\cos{\varphi}\sin{\vartheta},\\
\varrho_{2} & = & \varrho_{0}\sin{\varphi}\sin{\vartheta},\\
\varrho_{3} & = & \varrho_{0}\cos{\vartheta}.
\end{eqnarray}
We focus on the canonical part of the current densities $\tilde{\mathbf{J}}^{\prime}_{\mu}$ (the prime stands for the new gauge).
The Cartesian components of $\tilde{\mathbf{J}}^{\prime}_{\mu}$ are (for $k=x,y$)
\begin{eqnarray}
\tilde{J}_{0}^{{\prime}k} & = & \varrho_{0}\left.\big(1-\cos{\vartheta}\right.\big)\partial_{k}\varphi,\\
\tilde{J}_{1}^{{\prime}k} & = & \varrho_{0}\left.\big(\sin{\vartheta}\cos{\varphi}\partial_{k}\varphi+\sin{\varphi}\partial_{k}\vartheta\right.\big),\\
\tilde{J}_{2}^{{\prime}k} & = & \varrho_{0}\left.\big(\sin{\vartheta}\sin{\varphi}\partial_{k}\varphi-\cos{\varphi}\partial_{k}\vartheta\right.\big),\\
\tilde{J}_{3}^{{\prime}k} & = & -\tilde{J}_{0}^{k},\label{eq_J0J3}
\end{eqnarray}
The effective magnetic field $b_{\mu}=\frac{m_{e}}{\hbar}\nabla\times\mathbf{J}_{\mu}\cdot\hat{z}$
shall be shown to be responsible for the emergence of a Skyrmion in
the spin texture. Separating the canonical parts, the effective field
is written as
\begin{eqnarray}
b_{\mu}=\frac{1}{2}\tilde{b}_{\mu}-\frac{e}{\hbar}\left(\partial_{x}\varrho_{\mu}A_{y}^{\prime}-\partial_{y}\varrho_{\mu}A_{x}^{\prime}+B_{z}\varrho_{\mu}\right),
\end{eqnarray}
where $B_{z}=\nabla\times\mathbf{A}^{\prime}\cdot\hat{z}$ and
\begin{eqnarray}
\tilde{b}_{\mu}=\partial_{x}\tilde{J}_{\mu}^{{\prime}y}-\partial_{y}\tilde{J}_{\mu}^{{\prime}x}.
\end{eqnarray}
It will be shown that only the canonical parts contribute to the topological
density. The explicit form of $\tilde{b}_{\mu}$ for $\mu=0$ up to
$4$ are (summation over repeated indices for $p,q=x,y$ is implied)
\begin{eqnarray}
 &  & \tilde{b}_{0}=\left(1-\cos{\vartheta}\right)\epsilon_{pq}\partial_{p}\varrho_{0}\partial_{q}\varphi+\varrho_{0}\sin{\vartheta}\epsilon_{pq}\partial_{p}\vartheta\partial_{q}\varphi,\,\,\,\,\,\,\,\,\,\,\,\,\,\,\,\,\\
 &  & \tilde{b}_{1}=\sin{\vartheta}\cos{\varphi}\epsilon_{pq}\partial_{p}\varrho_{0}\partial_{q}\varphi+\sin{\varphi}\epsilon_{pq}\partial_{p}\varrho_{0}\partial_{q}\vartheta\nonumber \\
 &  & \,\,\,\,\,\,\,\,\,\,\,\,-\varrho_{0}\cos{\varphi}(1-\cos{\vartheta})\epsilon_{pq}\partial_{p}\vartheta\partial_{q}\varphi,\\
 &  & \tilde{b}_{2}=\sin{\vartheta}\sin{\varphi}\epsilon_{pq}\partial_{p}\varrho_{0}\partial_{q}\varphi-\cos{\varphi}\epsilon_{pq}\partial_{p}\varrho_{0}\partial_{q}\vartheta\nonumber \\
 &  & \,\,\,\,\,\,\,\,\,\,\,\,-\varrho_{0}\sin{\varphi}(1-\cos{\vartheta})\epsilon_{pq}\partial_{p}\vartheta\partial_{q}\varphi\\
 &  & \tilde{b}_{3}=-\tilde{b}_{0},
\end{eqnarray}
where $\epsilon_{xy}=-\epsilon_{yx}=1$ and zero otherwise. The following
identity
\begin{eqnarray}
\frac{\varrho_{\nu}}{\varrho_{0}}\eta^{\nu\mu}\frac{b_{\mu}}{\varrho_{0}}=\frac{1}{2}\frac{\varrho_{\nu}}{\varrho_{0}}\eta^{\nu\mu}\frac{\tilde{b}_{\mu}}{\varrho_{0}},
\end{eqnarray}
is held since $\mathbf{A}^{\prime}\varrho_{\nu}\eta^{\nu\mu}\varrho_{\mu}=0$
and hence its curl also vanishes. It is straightforward to show the
last identity gives the topological density in eq.~\eqref{eq_Q_02}
\begin{eqnarray}
J_{t}=\frac{\varrho_{\nu}}{\varrho_{0}}\eta^{\nu\mu}\frac{b_{\mu}}{\varrho_{0}}=\sin{\vartheta}\epsilon_{pq}\partial_{p}\vartheta\partial_{q}\varphi,
\end{eqnarray}
which is the Mermin-Ho relation \cite{Skyrmion}.

\section{Exact relation between spin fields and in-plane magnetic field}
\label{AppExRe}
We show that if $H$ does not give a good spin quantum number but it
has a rotational symmetry $[H,\jmath^{N}_{z}]=0$, where
$\jmath^{N}_{z}$ is defined by $\jmath^{N}_{z}
	=
	-i\partial_{\theta}/N
	+
	\sigma_{z}/2$, for $N \in \mathbb{Z}$ and $N \neq 0$,
then $\varrho_{1} B_{2}=\varrho_{2} B_{1}$ is held exactly. The eigenfunctions and eigenenergy of $\jmath^{N}_{z}$ are
\begin{eqnarray}
	\jmath^{N}_{z}
	\begin{pmatrix}
      \mathcal{N}_{1} e^{im \theta}    \\
      \mathcal{N}_{2} e^{i(m+N) \theta}
	\end{pmatrix}
	=(\frac{m}{N}+\frac{1}{2})
	\begin{pmatrix}
      \mathcal{N}_{1} e^{im \theta}    \\
      \mathcal{N}_{2} e^{i(m+N) \theta}
	\end{pmatrix}
	,
\end{eqnarray}
for $m \in \mathbb{Z}$ and $\mathcal{N}_{1,2}$ are normalization constant which satisfy $|\mathcal{N}_{1}|^{2} +|\mathcal{N}_{1}|^{2}=1$.

We choose the generic eigenfunction of $H$ in a gauge given in Eq. (\ref{eq_FigFNG}), and hence, the vector potential is $\mathbf{A}^{\prime}=\mathbf{A}-(\hbar/e)\nabla S_{1}$. In this representation we found $\mathbf{\tilde{J}}^{{\prime}}_{0}=-\mathbf{\tilde{J}}^{\prime}_{3}$, see Eq. (\ref{eq_J0J3}), where $\mathbf{\tilde{J}}^{\prime}_{\nu}$ is the canonical part of the current density $\mathbf{J}_{\nu}$. The stationary continuity equations involving $\mathbf{J}_{0,3}$ then read
\begin{eqnarray}
	\frac{\hbar}{2m_{e}}\nabla \cdot \mathbf{\tilde{J}}^{\prime}_{0}-\frac{e}{m_{e}} \mathbf{A}^{\prime} \cdot \nabla \varrho_{0}&=&0,\\
	-\frac{\hbar}{2m_{e}}\nabla \cdot \mathbf{\tilde{J}}^{\prime}_{0}-\frac{e}{m_{e}} \mathbf{A}^{\prime} \cdot \nabla\varrho_{3}
	&=&\frac{\Delta}{\hbar}\left[\mathbf{B}\times\boldsymbol{\varrho}(\mathbf{r})\right]\cdot\hat{x}_{3}.\,\,\,\,\,\,
\end{eqnarray}
Summing them up, we find
\begin{eqnarray}
	-\frac{e}{m_{e}} \mathbf{A}^{\prime} \cdot \nabla \left( \varrho_{0}+ \varrho_{3}\right)=\frac{\Delta}{\hbar}\left[\mathbf{B}\times\boldsymbol{\varrho}(\mathbf{r})\right]\cdot\hat{x}_{3}.
\end{eqnarray}
If $H$ and $\jmath^{N}_{z}$ commutes, they can have a simultaneous eigenfunction which implies the problem is separable $\nabla \psi_{1,2} || \hat{\rho}$ and $\nabla S_{1,2} ||\hat{\theta}$. Given $[H,\jmath^{N}_{z}]=0$ implies also the vector potential $\mathbf{A}$ is rotationally invariant, we have $\mathbf{A}^{\prime} \cdot \nabla \left( \varrho_{0}+ \varrho_{3}\right)=0$ which shows that the relation $\varrho_2/\varrho_1 = B_2/B1$ is exact.

\section{A perturbation theory}

\label{AppPer} In this section, we derive the spin texture of a confined
2D electron in the presence of a homogeneous magnetic field and treat
the in-plane magnetic field $H^{\prime}=(\Delta/2)B_{0}b\rho\cos{(\theta)}\sigma_{x}=(\Delta/2)B_{0}bx\sigma_{x}$
within a perturbation theory. The unperturbed Hamiltonian is
\begin{eqnarray}
H_{0}=\frac{(\mathbf{p}-e\mathbf{A})^{2}}{2m_{e}}+\frac{1}{2}m_{e}\omega^{2}\rho^{2}+\frac{\Delta B_{0}}{2}\sigma_{z}.
\end{eqnarray}
where $\mathbf{A}=(B_{0}/2)(-y,x)$, the electron charge $e<0$, $\omega$
is the confinement frequency, $\Delta$ is a coupling factor, the
electron g factor is $g_{e}$, $\mu_{B}$ is the Bohr magneton, $B_{0}$
is a constant with a dimension of magnetic field, and $b$ has a dimension
of inverse length. Eigenfunctions of $H_{0}$ are
\begin{eqnarray}
\Phi_{n_{+}n_{-}}^{\sigma}=R_{n_{-}}^{n_{+}}e^{i(n_{-}-n_{+})\theta}|\sigma\rangle,
\end{eqnarray}
where $\sigma=\pm1$ is the spin eigenvalue, $n_{\pm}\in\mathbb{Z}^{+}\bigcap\{0\}$
are the principal quantum numbers, and the radial part of the eigenfunction
is
\begin{eqnarray}
R_{n_{-}}^{n_{+}}=\mathcal{N}_{n_{-}}^{n_{+}}e^{-\rho^{2}/2\ell^{2}}\left(\frac{\rho}{\ell}\right)^{n_{-}-n_{+}}L_{n_{+}}^{n_{-}-n_{+}}\left(\frac{\rho^{2}}{\ell^{2}}\right),
\end{eqnarray}
where $L_{\lambda}^{\kappa}(x)$ is the associated Laguerre polynomial, $\ell=\sqrt{\hbar/(m_{e} \Omega)}$, $\Omega=\sqrt{\omega^{2}+\omega_{c}^{2}}$ and $\omega_{c}=|e|B_{0}/(2m_{e})$.
The normalization factor is
\begin{eqnarray}
\mathcal{N}_{n_{-}}^{n_{+}}=\frac{(-1)^{n_{+}}}{\sqrt{\pi}\ell}\sqrt{\frac{n_{+}!}{n_{-}!}}.
\end{eqnarray}
The corresponding eigenenergies are
\begin{eqnarray}
E_{n_{+}n_{-}}^{\sigma}=\sum_{k=\pm}\hbar\Omega_{k}\left(n_{k}+\frac{1}{2}\right)+\frac{\Delta}{2}\sigma,
\end{eqnarray}
where $\Omega_{\pm}=\Omega\pm\omega_{c}$. Treating $H^{\prime}=(\Delta/2)B_{0}b x \sigma_{x}$
in a first order perturbation theory, the eigenfunction $\Phi_{n_{+}n_{-}}^{\sigma}$
is modified to
\begin{eqnarray}
\widetilde{\Phi}_{n_{+}n_{-}}^{\sigma}=\Phi_{n_{+}n_{-}}^{\sigma}+\frac{1}{2}
\sum_{k_{+},k_{-}}\alpha_{k_{-}}^{k_{+}}\Phi_{k_{+}k_{-}}^{-\sigma},
\end{eqnarray}
while the non-zero coefficients are
\begin{eqnarray}
\alpha_{n_{-}}^{n_{+}+1} & = & \frac{\Delta B_{0}b\ell\sqrt{n_{+}+1}}{\Delta B_{0}\sigma-\hbar\Omega_{+}},\\
\alpha_{n_{-}}^{n_{+}-1} & = & \frac{\Delta B_{0}b\ell\sqrt{n_{+}}}{\Delta B_{0}\sigma+\hbar\Omega_{+}},\\
\alpha_{n_{-}+1}^{n_{+}} & = & \frac{\Delta B_{0}b\ell\sqrt{n_{-}+1}}{\Delta B_{0}\sigma-\hbar\Omega_{-}},\\
\alpha_{n_{-}-1}^{n_{+}} & = & \frac{\Delta B_{0}b\ell\sqrt{n_{+}+1}}{\Delta B_{0}\sigma+\hbar\Omega_{-}}.
\end{eqnarray}
The induced spin fields via the perturbation are
\begin{eqnarray}
\varrho_{1} & = & F_{1}^{n_{+}n_{-}}\cos{(\theta)},
%
%
\\
\varrho_{2} & = & \sigma F_{2}^{n_{+}n_{-}}\sin{(\theta)},
%
%
\end{eqnarray}
with the auxiliary functions
\begin{eqnarray*}
 &  & F_{1}^{n_{+}n_{-}}=R_{n_{-}}^{n_{+}}%
\sum_{k_{+},k_{-}}\alpha_{k_{-}}^{k_{+}}R_{k_{-}}^{k_{+}},%
\\
 &  & F_{2}^{n_{+}n_{-}}=R_{n_{-}}^{n_{+}}%
\sum_{k_{+},k_{-}}(k_{-}-k_{+}-n_{-}+n_{+})\alpha_{k_{-}}^{k_{+}}R_{k_{-}}^{k_{+}}.%
\end{eqnarray*}
If $\Delta B_{0}<\hbar\Omega_{\pm}$, then the states with the same
$n_{\pm}$ but opposite spin quantum numbers have winding numbers
which differ in sign $q=\pm1$ in the presence of the perturbation.

\section{Spin-orbit couplings}\label{APPSOC}

The spin texture of confined electrons in a 2D space subject to Rashba
or Dresselhaus spin-orbit coupling (SOC) hosts a spin vortex with
$|q|=1$ \cite{SOCST,Wenchen}. We use Eq.~\eqref{eqPhaB02} to count the
winding number of the vortices. The following Hamiltonian describes
a 2D electron in the presence of SOCs and a homogeneous magnetic field
\begin{eqnarray}
H_{SOC} & = & H_{0}+g_{1}\left(\left[p_{y}-eA_{y}\right]\sigma_{x}-\left[p_{x}-eA_{x}\right]\sigma_{y}\right)\\
 & + & g_{2}\left(\left[p_{y}-eA_{y}\right]\sigma_{y}-\left[p_{x}-eA_{x}\right]\sigma_{x}\right)+\frac{\Delta}{2}B_0\sigma_{z},\nonumber
\end{eqnarray}
where $H_{0}$ is given in Eq.~\eqref{eq_H0}, $\mathbf{A}=(A_{x},A_{y},0)=(B_{0}/2)(-y,x,0)$
is the vector potential, the strength of Rashba (Dresselhaus) term
is determined by $g_{1}$ $(g_{2})$, and $\Delta$ is a coupling
constant. Due to the broken translational symmetry, the electron has
confinement lengths $L_{x,y}$. Therefore, the density of momentum
decays to zero for distances larger than the confinement lengths $\rho\gtrsim L_{x,y}$
and can be neglected in comparison to the other terms in the Hamiltonian.
Neglecting the derivative terms in $H_{SOC}$ in the asymptotic regime,
we have
\begin{eqnarray}
\widetilde{H}_{SOC} & = & \widetilde{H}_{0}-eg_{1}\left(A_{y}\sigma_{x}-A_{x}\sigma_{y}\right)\\
 & - & eg_{2}\left(A_{y}\sigma_{y}-A_{x}\sigma_{x}\right)+\frac{\Delta}{2}B_0\sigma_{z},\nonumber
\end{eqnarray}
where $\widetilde{H}_{0}$ also does not include derivative terms.
The effective magnetic fields at $\rho\gg L_{x,y}$ associated with
the SOCs are then
\begin{eqnarray}
\mathbf{B}_{R} & = & -eg_{1}\left(A_{y},-A_{x},0\right),\\
\mathbf{B}_{D} & = & -eg_{2}\left(-A_{x},A_{y},0\right).\nonumber
\end{eqnarray}
Putting these effective fields into Eq.~\eqref{eqPhaB02}, the winding
number for the Rashba-dominant SOC (i.e. $g_{1}>g_{2}$) is $q_{R}=+1$,
for the case of Dresselhaus-dominant SOC (i.e. $g_{1}<g_{2}$) is
$q_{D}=-1$, and for $g_{1}=g_{2}$ the winding number vanishes. Since
the spin densities are observables, the results are independent of
the gauge of the vector potential.


\begin{thebibliography}{10}
 \bibitem{Nat547} Davide Castelvecchi, Nature \textbf{547}, 272-274 (2017).
 \bibitem{thoul} D. J. Thouless, \textit{Topological Quantum Numbers in Nonrelativistic Physics} (World Scientific, River Edge, N.J., 1998).
 \bibitem{berry01} M. V. Berry, Proc. R. Soc. London A \textbf{392}, 45 (1984).
 \bibitem{berry02}  D. Xiao, M.-C. Chang, and Q. Niu, Rev. Mod. Phys. \textbf{82}, 1959 (2010).

 \bibitem{dot01} Yu. V. Nazarov and Ya. M. Blanter, Quantum Transport: Introduction to Nanoscience (Cambridge University Press, Cambrige, England, 2009).
 \bibitem{dot02} S. M. Reimann and M. Manninen, Rev. Mod. Phys. 74, 1283 (2002).

 \bibitem{PRA57} D. Loss and D. P. DiVincenzo, Phys. Rev. A \textbf{57}, 120 (1998).


 \bibitem{Nat489} J. J. Pla, K. Y. Tan, J. P. Dehollain, W. H. Lim, J. J. L. Morton, D. N. Jamieson, A. S. Dzurak, and A. Morello,


  Nature \textbf{489}, 541 (2012).

\bibitem{Zutic} I. Zutic, J. Fabian, and S. Das Sarma, Rev. Mod. Phys. \textbf{76}, 323 (2004).
\bibitem{Smejkal} L. Smejkal, Y. Mokrousov, Binghai Yan, and A. H. MacDonald, Nat. Phys. \textbf{14}, 242 (2018).
\bibitem{Sinova} O. Gomonay, T. Jungwirth, and J. Sinova, Phys. Stat. Sol. RRL \textbf{11}, 1700022 (2017).


\bibitem{science339} David D. Awschalom, Lee C. Bassett, Andrew S. Dzurak, Evelyn L. Hu, Jason R. Petta, Science \textbf{339}, 6124, (2013).
\bibitem{science318} K. C. Nowack, F. H. L. Koppens, Yu. V. Nazarov, L. M. K. Vandersypen, Science \textbf{318}, 5855, (2007).
 \bibitem{PRL122} L. C. Camenzind, L. Yu, P. Stano, Zimmerman, A. C. Gossard, D. Loss, and D. M. Zumb\"uhl,
 Phys. Rev. Lett. \textbf{122}, 207701 (2019).
 \bibitem{PRB99} P. Stano, C.-H. Hsu, L. C. Camenzind, L. Yu, Dominik Zumb\"uhl, and D. Loss,
  Phys. Rev. B \textbf{99}, 085308 (2019).
 \bibitem{SOCST} Wenchen Luo, Amin Naseri, Jesko Sirker, Tapash Chakraborty, Sci. Rep. {\bf 9}, 672 (2019).

 \bibitem{elvo01} K. Y. Bliokh, I. P. Ivanov, G. Guzzinati, L. Clark, R. Van Boxem, A. B\'ech\'e, R. Juchtmans, M. A. Alonso, P. Schattschneider, F. Nori, J.Verbeeck, Phys. Rep. \textbf{690} (2017).
 \bibitem{elvo02} S. M. Lloyd, M. Babiker, G. Thirunavukkarasu, and J. Yuan, Rev. Mod. Phys. \textbf{89}, 035004 (2017).
 \bibitem{elvo03} K. Y. Bliokh, Y. P. Bliokh, S. Savel'ev, and F. Nori, Phys. Rev. Lett. \textbf{99}, 190404 (2007).
 \bibitem{elvo04} M. Uchida and A. Tonomura, Nature \textbf{464}, 737 (2010).
 \bibitem{elvo05} E. Karimi, L. Marrucci, V. Grillo, and E. Santamato, Phys. Rev. Lett. \textbf{108}, 044801 (2012).
 \bibitem{Nogaret} A. Nogaret, J. Phys. Condens. Matter \textbf{22},  253201 (2010).
 \bibitem{Kondo} A. C. Hewson, \textit{The Kondo Problem to Heavy Fermions} (Cambridge University Press, Cambridge, U.K., 1993).
 \bibitem{Skyrmion} Jung Hoon Han, \textit{Skyrmions in Condensed Matter}, (Springer International Publishing, 2017).
 \bibitem{Nakahara} M. Nakahara, \textit{Geometry, Topology and Physics} (IOP, Bristol, 1990).
 \bibitem{spcu01} Junren Shi, Ping Zhang, Di Xiao, and Qian Niu, Phys. Rev. Lett. \textbf{96}, 076604 (2006).
 \bibitem{spcu02} Z. An, F. Q. Liu, Y. Lin and C. Liu, Sci. Rep. \textbf{2}, 388 (2012).
 \bibitem{inho01} F. M. Peeters and A. Matulis, Phys. Rev. B \textbf{48}, 15166 (1993).
  \bibitem{Wenchen} Wenchen Luo and Tapash Chakraborty, Phys. Rev. B {\bf 100}, 085309 (2019).



 \end{thebibliography}
\end{document}